\documentclass[preprint,review,12pt,sort&compress]{elsarticle}
\usepackage[top=1in,bottom=1in,left=1in,right=1in]{geometry}
\usepackage{longtable}
\usepackage{framed} 
\usepackage{multicol} 
\usepackage{nomencl} 
\makenomenclature
\usepackage{ifthen}
  \renewcommand{\nomgroup}[1]{%
  \item[\bfseries
  \ifthenelse{\equal{#1}{A}}{Abbreviation Acronyms}{%
  \ifthenelse{\equal{#1}{S}}{Symbols}{%
  \ifthenelse{\equal{#1}{N}}{Number sets}{}}}%
  ]}
\renewcommand*\nompreamble{\begin{multicols}{2}}
\renewcommand*\nompostamble{\end{multicols}}

\usepackage{amssymb}
\usepackage{amsmath}
\usepackage{siunitx}

\journal{Renewable Energy}

\begin{document}

\begin{frontmatter}



\title{On long-term fatigue damage estimation for a floating offshore wind turbine using a surrogate model}


\author[inst1]{Ding Peng Liu}
\ead{dpl@utexas.edu}

\affiliation[inst1]{organization={Dept. of Civil, Arch., and Environmental Engineering, The University of Texas at Austin, Texas, USA}}

\author[inst2]{Giulio Ferri}
\ead{giulio.ferri@unifi.it}
\author[inst1]{Taemin Heo}
\ead{taemin@utexas.edu}
\author[inst2]{Enzo Marino}
\ead{enzo.marino@unifi.it}
\author[inst1]{Lance Manuel}
\ead{lance.manuel@utexas.edu}
\affiliation[inst2]{organization={Dept. of Civil and Environmental Engineering, University of Florence},
            city={Florence},
            country={Italy}}

\begin{abstract}
This study is concerned with the estimation of long-term fatigue damage for a floating offshore wind turbine. With the ultimate goal of efficient evaluation of fatigue limit states for floating offshore wind turbine systems, a detailed computational framework is introduced and used to develop a surrogate model using Gaussian process regression. The surrogate model, at first, relies only on a small subset of representative sea states and, then, is supplemented by the evaluation of additional sea states that leads to efficient convergence and accurate prediction of fatigue damage. A 5-MW offshore wind turbine supported by a semi-submersible floating platform is selected to demonstrate the proposed framework. The fore-aft bending moment at the turbine tower base and the fairlead tension in the windward mooring line are used for evaluation. Metocean data provide information on joint statistics of the wind and wave along with their relative likelihoods for the installation site in the Mediterranean Sea, near the coast of Sicily.  \textcolor{black}{A coupled frequency-domain model} provides needed power spectra for the desired response processes. The proposed approach offers an efficient and accurate alternative to the exhaustive evaluation of a larger number of sea states and, as such, avoids excessive response simulations.
\end{abstract}




\end{frontmatter}


\begin{table*}
\begin{framed}
\tiny
\nomenclature[A1]{DoF}{Degree of freedom}
\nomenclature[A2]{FOWT}{Floating offshore wind turbine}
\nomenclature[A3]{FD}{Frequency domain}
\nomenclature[A4]{GPR}{Gaussian process regression}
\nomenclature[A5]{MCS}{Monte Carlo simulation}
\nomenclature[A6]{MSL}{Mean sea level}
\nomenclature[A7]{PDF}{Probability density function}
\nomenclature[A8]{PSD}{Power spectral density}

\nomenclature[S10]{\(a\)}{Amplitude of the heave plates' oscillations}
\nomenclature[S11]{\(A_{Moor}\)}{Mooring line cross-sectional area}
\nomenclature[S12]{\(\textbf{A}(\omega\))}{Frequency-dependent hydrodynamic added mass matrix}
\nomenclature[S13]{\(b\)}{Inverse slope for log-scale S-N curve}
\nomenclature[S14]{\(\textbf{B}_M\)(\(\omega\),\(a\),\(\sigma\))}{Viscous drag damping matrix}
\nomenclature[S15]{\(\textbf{B}_T\)}{Damping matrix of the wind turbine}
\nomenclature[S16]{\(\textbf{B}(\omega\))}{Frequency-dependent hydrodynamic radiation damping matrix}
\nomenclature[S17]{\(\textbf{C}_H\)}{Hydrostatic stiffness matrix}
\nomenclature[S18]{\(CI(T,x^*)\)}{ Confidence interval for short-term damage}
\nomenclature[S19]{\(\textbf{C}_{Moor}\)}{Mooring system stiffness matrix}
\nomenclature[S20]{\(\textbf{C}_T\)}{Stiffness matrix of the wind turbine}
\nomenclature[S21]{\(\textbf{C}^{TOT}\)}{Total stiffness matrix}
\nomenclature[S22]{\(\hat D\)}{Predicted short-term fatigue damage}
\nomenclature[S23]{\(DEL_{1-Hz}\)}{1-Hz damage-equivalent load}
\nomenclature[S24]{\(D(T)\)}{ Short-term fatigue damage experienced over time}
\nomenclature[S25]{\(\textbf{F}_d\)}{Modal aerodynamic transfer function}
\nomenclature[S26]{\(G(f)\)}{One-sided power spectral density function}
\nomenclature[S27]{\(GP(\mu,k)\)}{Gaussian process prior}
\nomenclature[S28]{\(G_{1},G_{2},G_{3},Q,R\)}{Parameters for Dirlik method}
\nomenclature[S29]{\(\textbf{h}_{TT}\)}{Height of the tower top}
\nomenclature[S30]{\(h_{1},h_{2}\)}{Smoothing parameters}
\nomenclature[S31]{\(H_{s}\)}{Significant wave height}
\nomenclature[S32]{\(\textbf{I}\)}{Identity matrix}
\nomenclature[S33]{\(k_{f|X,y}(x^*,x^{*'})\)}{Predictive posterior covariance}
\nomenclature[S34]{\(k(x,x')\)}{Radial basis function kernel}
\nomenclature[S35]{\(K_{a}\)}{Intercept of N in the S-N curve}
\nomenclature[S36]{\(Ker\)}{Selected non-negative Gaussian kernel function}
\nomenclature[S37]{\(\textbf{K}^p\)}{Single cable stiffness matrix}
\nomenclature[S38]{\(\textbf{K}_{XX}\)}{The matrix of all the kernel entry pairs}
\nomenclature[S39]{\(l\)}{Horizontal scale of radial basis function kernel}
\nomenclature[S40]{\(LTD(T)\)}{ Long-term fatigue damage experienced over time}
\nomenclature[S41]{\(L(\theta|X,y)\)}{ Log marginal likelihood function}
\nomenclature[S42]{\(\textbf{M}_F\)}{Floater mass matrix of the }
\nomenclature[S43]{\(m_{j}\)}{$j$th spectral moment}
\nomenclature[S44]{\(\textbf{M}_T\)}{Wind turbine mass matrix}
\nomenclature[S45]{\(M_{TWB}\)(\(\omega\))}{Bending moment at the tower base}
\nomenclature[S46]{\(n\)}{The number of wave data points in the wind speed bin}
\nomenclature[S47]{\(n(T)\)}{ Total number of cycles encountered over time}
\nomenclature[S48]{\(N(S)\)}{The number of cycles that lead to failure for stress amplitude}

\nomenclature[S49]{\(\hat p\)}{Joint probability density function of sea state}
\nomenclature[S50]{\(p(S)\)}{Probability density function of stress range amplitude}
\nomenclature[S51]{\(\textbf{q}_\text{F}\)(\(\omega\))}{In-plane cable fairlead displacement vector}
\nomenclature[S52]{\(\textbf{q}(\omega\))}{Frequency-dependent system response vector}
\nomenclature[S53]{\(q_\text{5}\)(\(\omega\))}{Pitch rigid body rotation of the system}
\nomenclature[S54]{\(q_\text{7}\)(\(\omega\))}{Tower-top fore-aft deflection}
\nomenclature[S55]{\(R_{q5}(\omega)\))}{Rotation transformation matrix associated with pitch motion}
\nomenclature[S56]{\(s\)}{Vertical scale of radial basis function kernel}
\nomenclature[S57]{\(S\)}{Stress range amplitude}
\nomenclature[S58]{\(S_{waves}(\omega)\)}{Power spectral density of wave elevation}
\nomenclature[S59]{\(S_{wind}(\omega)\)}{Power spectral density of wind velocity}
\nomenclature[S60]{\(T\)}{Exposure time}
\nomenclature[S61]{\(T_{F}\)(\(\omega\))}{Mooring line tension}
\nomenclature[S62]{\(T_{p}\)}{Wave peak period}
\nomenclature[S63]{\(U\)}{Mean wind speed at \SI{10}{\meter} above MSL}
\nomenclature[S64]{\(V\)}{Mean wind speed at hub height}
\nomenclature[S65]{\(V_{TWB}\)(\(\omega\))}{Shear force at the tower base}
\nomenclature[S66]{\(W_{TWB}\)}{Section modulus of the tower base}
\nomenclature[S67]{\(\textbf{x}^*\)}{Test point}
\nomenclature[S68]{\(X\)}{Training data input}
\nomenclature[S69]{\(\textbf{X}(\omega\))}{Frequency-dependent diffraction force matrix}
\nomenclature[S70]{\(\textbf{y}\)}{Training data output}
\nomenclature[S71]{\(Z\)}{Normalized stress amplitude}
\nomenclature[S72]{\(\gamma\)}{Standard score}
\nomenclature[S73]{\(\epsilon\)}{Gaussian noise}
\nomenclature[S74]{\(\theta\)}{Hyperparameters}
\nomenclature[S75]{\(\mu(x)\)}{Gaussian prior mean}
\nomenclature[S76]{\(\mu_{f|X,y}(x^*)\)}{Predictive posterior mean}
\nomenclature[S77]{\(\nu_p\)}{Peak occurrence rate}
\nomenclature[S78]{\(\sigma\)}{Standard deviation of the platform motion}
\nomenclature[S79]{\(\sigma(x^*)\)}{Standard deviation of the prediction}
\nomenclature[S80]{\(\sigma_{Fair}\)}{Normal stress at the fair lead}
\nomenclature[S81]{\(\sigma_{TWB}\)}{Bending stress at the tower base}
\printnomenclature
\end{framed}
\end{table*}

\newpage
\section{Introduction}
\label{sec:introduction}
Floating offshore wind turbines (FOWTs) are increasingly being considered in projects seeking to generate electricity from wind energy offshore, especially at deep water sites where the resource is often abundant and less influenced by the surface roughness at near-coast sites and at onshore installations~\cite{Xu_RSER2020}. Complex wind-wave-structure interaction, however, that can result both at rigid elements (the floating platform and nacelle) and at flexible elements (the tower, blades, and moorings), exposes the integrated system to dynamic loads that can lead to premature failure of important components, such as the mooring lines and the turbine tower~\cite{ThiesRE,HsuMS,Liu2108JSEE}. For this reason, the design of FOWT systems must consider and evaluate the possibility of long-term fatigue damage which relies on a full understanding of the actual joint metocean statistcs at the planned installation site and must consider relevant load cases so as to assess the reliability of the system over the long term against various limit states~\cite{manuel2018alternative,MarinoRE,ZieglerEP}.

For site-specific fatigue assessments, sea states at the offshore site of interest should be selected based on their likelihood, anticipated severity, and possible interaction with the floating system; such distinct sea states identifying combinations of wind speed, wave height and wave period are generally evaluated as part of the so-called ``short-term'' fatigue damage evaluation. Note that, given specific values of these variables, input wind- and wave-related standard power spectra can have additional parametric uncertainties that propagate directly to short-term turbine loads~\cite{JWEIA2008Korn}. 
In such short-term evaluations, the aero-hydro-servo-elastic time-domain simulations~\cite{jonkman2011dynamics} \textcolor{black}{or computational fluid dynamics (CFD) simulations that capture complex aerodynamic effect \cite{liu2017establishing}} can be computationally expensive~\cite{kvittem2015frequency}. Coupled system models formulated in the frequency domain (FD) have been proposed that offer an appealing alternative for reducing computational costs~\cite{karimi2017multi,karimi2019fully,Ferri2022,Ferri2023}.
A ``long-term'' fatigue damage assessment must then incorporate all such short-term evaluations in the overall long-term fatigue reliability assessment over the planned service life. 

\textcolor{black}{Computationally expensive CFD simulations can sometimes be required to estimate FOWT fatigue damage while accounting for complex dynamics such as due to unsteady aerodynamic effects associated with platform surge motion \cite{tran2015aerodynamic,kyle2020propeller} or pitch and yaw motions \cite{tran2015aerodynamic} as well as blade-tower interactions in aerodynamic loading~\cite{cai2023effects,dose2020fluid,santo2020effect}. For both CFD and coupled FD simulation cases, the computational cost incurred in estimating long-term fatigue damage can be reduced by developing and employing surrogate models in lieu of high-fidelity coupled models.} In the offshore wind industry and other ocean engineering applications, reduced-order surrogate models have been developed when dealing with time-consuming computations, such as with structural finite element models \cite{zheng2023efficient}, in CFD \cite{wilson2017surrogate,zahle2018computational,lim2021RESS}, and in condition monitoring \cite{li2019wind,avendano2021virtual,zhang2022hybrid}. In structural design, due to their established efficiency, surrogate models have been applied for both extreme loads analysis \cite{abdallah2019parametric,taflanidis2013offshore,lim2021JOMAE} and for fatigue reliability analysis \cite{leimeister2018review,wilkie2021gaussian,pillai2018mooring,Lim2022ES}. Regression models (using Gaussian process regression (GPR), polynomial chaos expansion, etc.)  \cite{singh2022probabilistic,teixeira2017analysis,murcia2018uncertainty,shi2019AOR} and machine learning approaches such as with artificial neural networks \cite{muller2018application,richmond2020stochastic},  have all been adopted to build such surrogate models. Among all these methods, GPR has proven to be an appropriate choice for describing fatigue loads  \cite{singh2022probabilistic,gasparis2020surrogate,li2020long} although it may sometimes require greater computational cost \cite{dimitrov2018wind}. In this study, we present an efficient fatigue damage assessment framework based on a GPR model that serves as a surrogate to estimate the short-term fatigue for input sea states.

In constructing and training surrogate models that must account for model uncertainty, a large training set may be required with GPR, in order to yield good predictions that describe the uncertainty \cite{ankenman2008stochastic,yue2018surrogate}, if such a set can be easily obtained. Instead of using all the available data to train a GPR model, selecting an appropriate (smaller) training data set is reasonable. For instance, iterative trimming of the outliers has been proposed to improve the performance of the surrogate model \cite{wang2017robust,li2021robust}. On the other hand, if the training data available are limited, efficiency and effective information acquisition become important. To achieve good GPR model accuracy using a small number of simulations, active learning with the GPR model has been proposed \cite{pasolli2011gaussian,yue2020active} that leverages both exploration and exploitation. Exploitation focuses on anomalous and interesting regions associated with high uncertainty while exploration, generally preceding exploitation, focuses on overall space-filling \cite{lam2008sequential,yan2017high}. In contrast with iterative trimming, in this latter approach, new data are readily and easily added to the existing training set and thus offer feasible exploitation routes  \cite{pasolli2011gaussian,kowalska2012maritime,yue2020active}. 

The approach we adopt in this study seeks to generate a response surface for damage-equivalent fatigue loads based on GPR and employs carefully selected sea states to make up the training data set. By introducing an active learning strategy, this approach ensures more accurate predictions with a smaller number of FD simulations than is possible with more conventional surrogate model approaches. In this manner, the efficiency of long-term fatigue assessment for the selected FOWT can be significantly improved.  In our study, first, a small subset of representative sea states is selected, following a metocean analysis, and then used to build an initial GPR surrogate model. This initial selection of sea states ensures good exploration of the variable space with consideration of the joint probability of variables defining the sea states. After simulations for these selected sea states using a verified FD model developed by Ferri et al.~\cite{Ferri2022,Ferri2023}, the relevant fatigue loads assessment is conducted using output load/response power spectra. Derived (output) damage-equivalent loads along with the (input) sea states make up the training data sets for the GPR surrogate model building. We show how improved response surfaces result from adding additional training data sets that consider the predictive uncertainty in short-term fatigue damage estimation; this enhancement represents the active learning associated with exploitation. Finally, long-term fatigue damage is then estimated using predictions based on the improved GPR-based response surfaces. \textcolor{black}{The experiment conducted on the OC4 DeepCwind 5-MW FOWT, installed in the Mediterranean Sea near the coast of Sicily, demonstrates that our GPR approach requires about ten times fewer simulations to estimate the long-term fatigue damage with an error margin of only 0.2\%. This remarkable efficiency suggests that our method facilitates faster estimation of long-term fatigue damage given a set amount of computational resources, thereby saving significant time and energy. Moreover, fast and accurate estimation of the long-term fatigue damage can enable comprehensive reliability-based design optimization for FOWTs \cite{stieng2020reliability}. Such optimization has the potential to reduce the levelized cost of energy by lowering construction costs and extending FOWT service life for energy generation.}

\section{Methodology}
\label{sec:Methodology}
\subsection{Problem Formulation}
The present study outlines a comprehensive framework that combines metocean data analysis; coupled FD response analysis of FOWTs; spectral fatigue damage assessment; and Gaussian process regression---all with the goal of efficient construction of a surrogate model for reliable and fast predictions of fatigue damage.

The long-term fatigue damage estimation procedure is demonstrated for the OC4 DeepCwind 5-MW FOWT, designed for a water depth of \SI{200}{\meter}. The support structure is a semi-submersible floater anchored to the seabed by three catenary mooring lines \textcolor{black}{(The middle image in Fig.~\ref{fig:FD} shows the structure of the platform)}. The platform has a draft of \SI{20}{\meter} and a freeboard of \SI{12}{\meter}. It has three side columns connected by means of cross braces and pontoons to a central column, which directly supports the wind turbine tower. Additionally, at the base of each side column, there is a heave plate with thickness and diameter equal to \SI{6}{\meter} and \SI{24}{\meter}, respectively. The mooring lines are spread symmetrically about the platform's vertical axis, with one of the cables aligned with the 0-degree wind heading direction; the lines are comprised of steel chain with a diameter of \SI{0.0766}{\meter}. The wind turbine tower base is connected to the platform at an elevation of \SI{10}{\meter} above the mean sea level (MSL), while the tower top is \SI{87.6}{\meter} above the MSL; this implies a flexible tower height of \SI{77.6}{\meter}. Both the diameter and the wall thickness of the tower vary linearly with height from \SI{6.5}{\meter} and \SI{0.027}{\meter} at the base to \SI{3.87}{\meter} and \SI{0.019}{\meter}, respectively, at the top. Additional details related to the structure can be found in \cite{oc42014}.

\subsection{Coupled Simulations in the Frequency Domain} 
\begin{figure}[!htb]
\unskip
\centering
\includegraphics[width =0.55\textwidth]{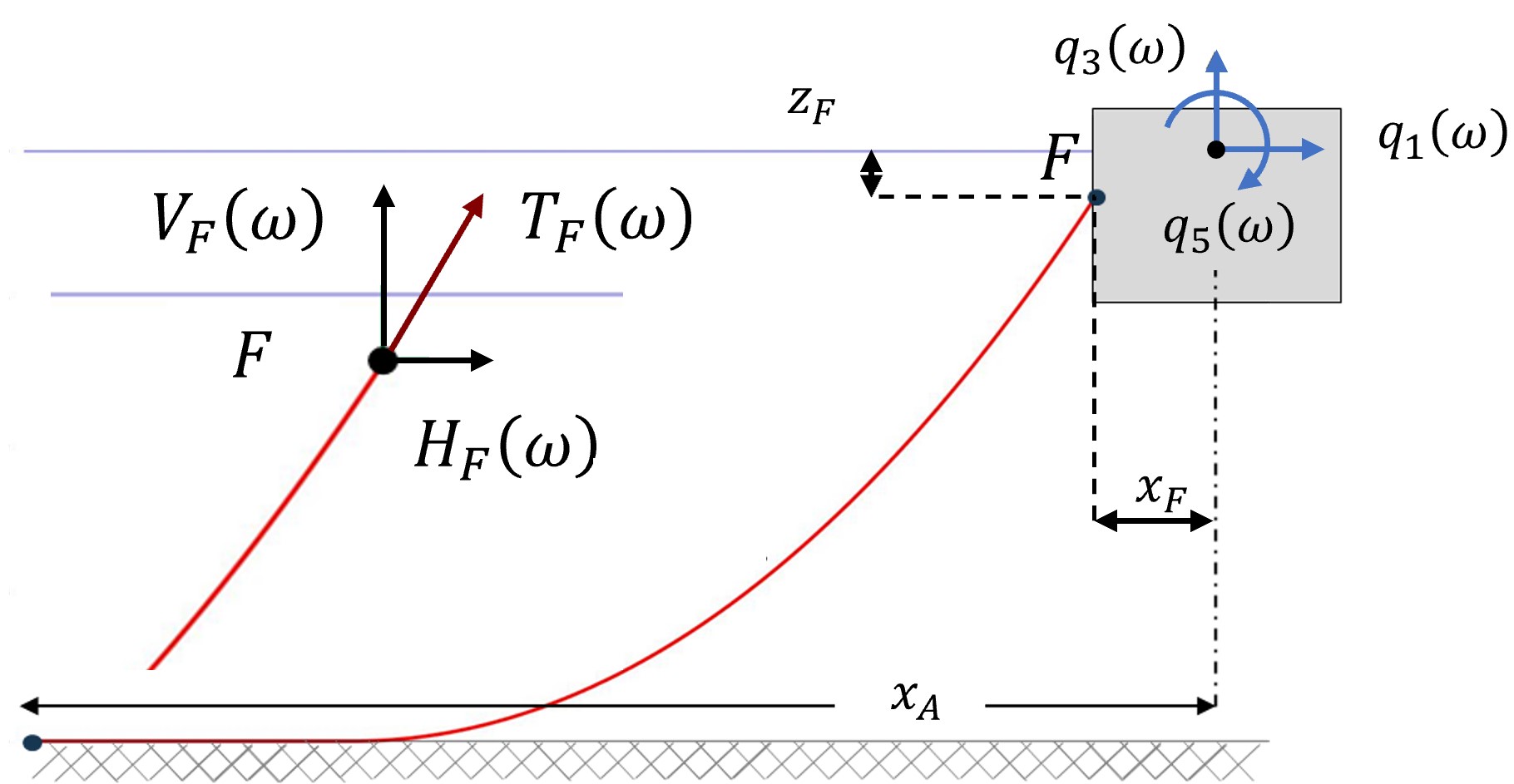}
\caption{Computation of the mooring line tension, $ T_F(\omega)$, from the rigid motions of the fairlead point, $\mathbf{q_{F}}(\omega)$.
\label{fig:normal_moor}}
\end{figure}

\begin{figure}[!htb]
\unskip
\centering
\includegraphics[width =0.3\textwidth]{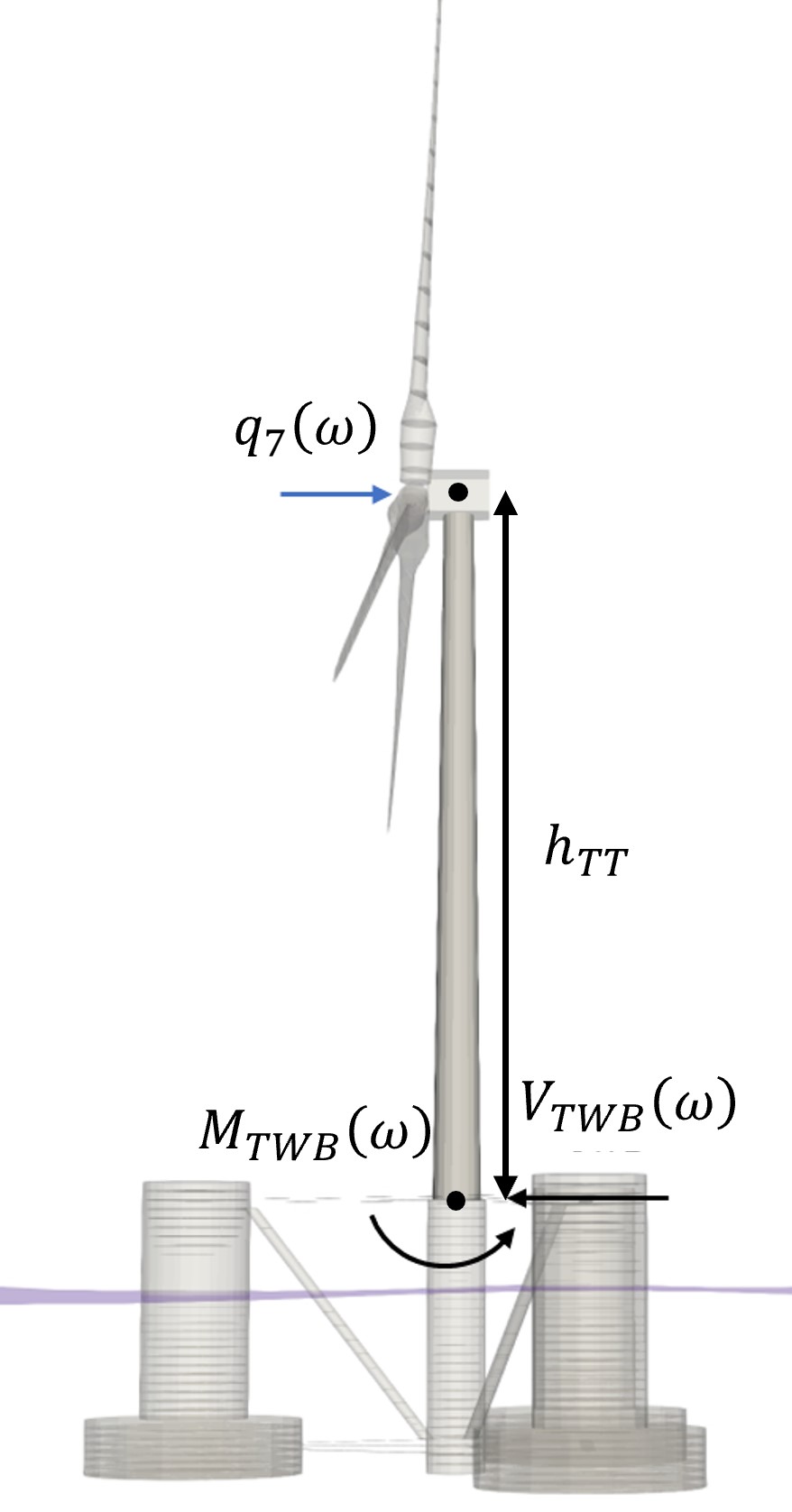}
\caption{Computation of the tower base bending moment, $ M_{TWB}(\omega)$, from the elastic deflection of the tower, $q_7(\omega)$.
\label{fig:normal_tower}}
\end{figure}

\begin{figure}[!htb]
\unskip
\centering
\includegraphics[width = 1\textwidth]{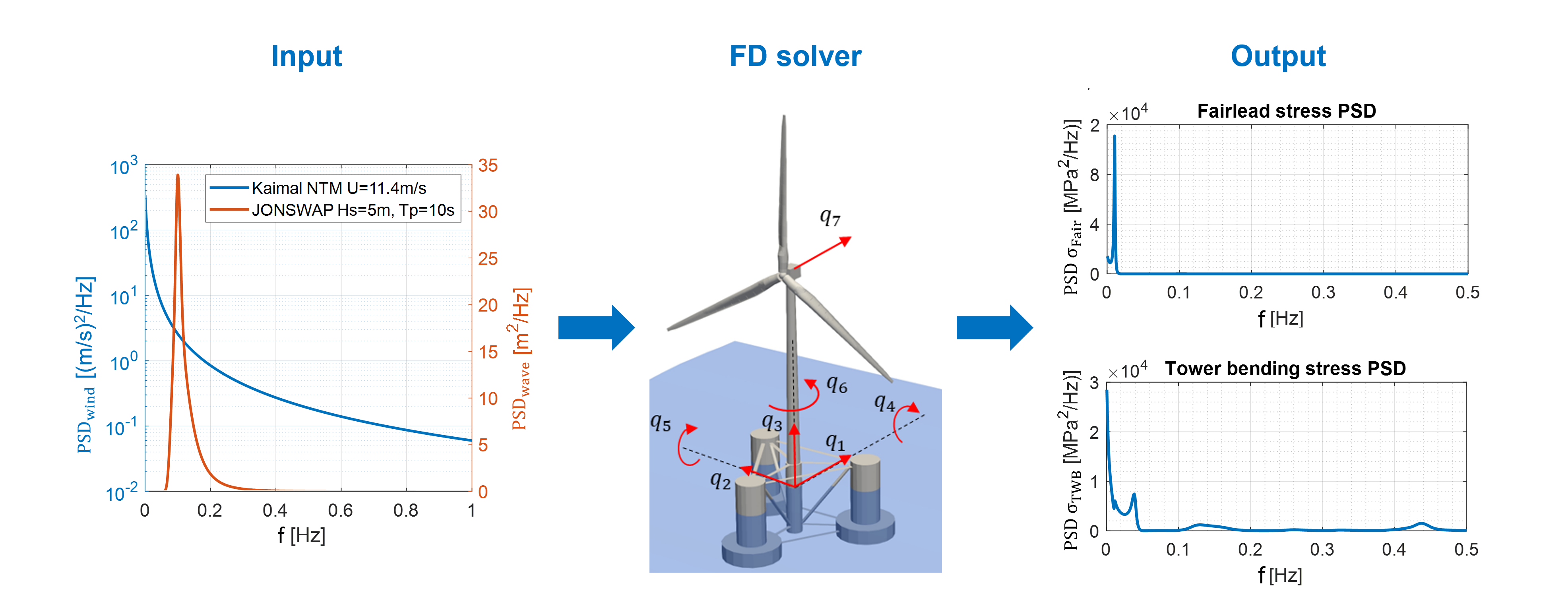}
\caption{Frequency domain framework for the calculation of stresses. 
\label{fig:FD}}
\end{figure}

Response simulations for a selected set of sea states are performed by employing a coupled FD formulation, described in detail in~\cite{Ferri2023}. \textcolor{black}{The FOWT is modeled as a system with 7 degrees of freedom (DoFs), as shown in Fig.~\ref{fig:FD}, where 6 DoFs are related to the floater's rigid-body motions and one accounts for the wind turbine's tower-top deflection in the fore-aft direction.} Hydrodynamic contributions stemming from the first-order wave-structure interaction are modeled considering both potential flow theory and Morison's equation. The damping contribution associated with the drag term in Morison's equation is linearized according to~\cite{Borgman1969}. The wind turbine and mooring line effects on the dynamics of the 7-DoF system are accounted for in FAST simulations~\cite{Jonkman2007},  linearized around a steady-state operating point. Loads are defined by means of power spectra, adopting Kaimal spectra for the turbulent wind and JONSWAP spectra for the waves. The governing equations of motion of the 7-DoF coupled system can be represented as:
\begin{equation}
\begin{split}
&[-\omega^2(\mathbf{A}(\omega)+\mathbf{M}_{F}+\mathbf{M}_{T})+i\omega(\mathbf{B}(\omega)+\mathbf{B}_{T}+\mathbf{B}_M(\omega,a,\sigma))+\\
&(\mathbf{C}_{H}+\mathbf{C}_{T}+\mathbf{C}_{Moor})]\cdot\mathbf{q}(\omega) =\mathbf{X}(\omega)\sqrt{2S_{waves}(\omega)\Delta\omega}+\mathbf{F}_{d}\sqrt{S_{wind}(\omega)\Delta\omega}
\label{eq:eq1}
\end{split}
\end{equation}
where $\textbf{q}(\omega)$ describes the system response in terms of a vector comprised of the amplitudes of motion for the 7 DoFs; $\textbf{A}(\omega)$, $\textbf{B}(\omega)$, and $\textbf{X}(\omega)$ refer to the frequency-dependent hydrodynamic added mass, radiation damping and diffraction force matrices, respectively; $\textbf{M}_{F}$ is the mass matrix of the floater, calculated with respect to the MSL; $\textbf{M}_{T}$, $\textbf{B}_{T}$, and $\textbf{C}_{T}$ are, respectively, the mass, damping and stiffness matrices of the wind turbine, accounting for the elasticity of the tower and blades as well as gyroscopic effects arising from interaction of the rotating blades with motions of the floating platform; $\textbf{F}_{d}$ is a modal aerodynamic transfer function, calculated using FAST by employing a multi-blade coordinate transformation~\cite{Bir2008}; $\mathbf{B}_M(\omega,a,\sigma)$ is the viscous drag damping matrix, evaluated by employing the Borgman linearization~\cite{Borgman1969} and following the procedure presented in~\cite{Ferri2022} (this damping depends on the amplitude, $a$, of the heave plates' oscillations and on the standard deviation of the platform motion, $\sigma$; it is computed by means of an iterative procedure); $\mathbf{C}_{H}$ and $\mathbf{C}_{Moor}$ are the hydrostatic and mooring system stiffness matrices, respectively; $S_{waves}(\omega)$ and $S_{wind}(\omega)$ are the power spectral densities (PSDs) for the wave elevation and the  wind velocity, respectively.  A long-crested sea is assumed with the waves and wind aligned.
Solving the system of equations, Eq.~\eqref{eq:eq1}, for the selected sea states, power spectra of the normal stresses in the mooring lines (Fig.~\ref{fig:normal_moor}) lead to estimates of $\sigma_{Fair}$ at the fairlead and $\sigma_{TWB}$ at the tower base (Fig.~\ref{fig:normal_tower}) as follows: 
\begin{align}
\mathbf{q_{F}}(\omega) & = \begin{bmatrix}
q_1(\omega)\\
q_3(\omega)\\
\end{bmatrix}+(\mathbf{R}_{q_{5}}(\omega)-\mathbf{I}) \begin{bmatrix}
x_F\\
z_F\\
\end{bmatrix}\\
\begin{bmatrix}
H_F(\omega)\\
V_F(\omega)\\
\end{bmatrix} &=\mathbf{K}^{p}\cdot\mathbf{q_{F}}(\omega)\rightarrow T_F(\omega)= || \begin{bmatrix}
H_F(\omega) &
V_F(\omega)\\
\end{bmatrix}^{T} || \\
\sigma_{Fair}(\omega) &=\frac{T_F(\omega)}{A_{moor}}
\end{align}
where $\mathbf{q_{F}}(\omega)$ is the in-plane cable fairlead displacement vector; $\mathbf{R}_{q_{5}}(\omega)$ is a rotation transformation matrix associated with pitch rigid body rotation of the system, ${q_{5}}(\omega)$, $\mathbf{I}$ is the identity matrix, while $\mathbf{K}^{p}$ is the 2x2 stiffness matrix of the single cable with respect to the fairlead displacement \cite{Ferri2022}; and $A_{moor}$ is the mooring line cross-sectional area.

At the tower base, we have:
\begin{align}
M_{TWB}(\omega) & = V_{TWB}(\omega)h_{TT}=[\mathbf{C}^{TOT}(7,7)q_7(\omega)]h_{TT}\\
\sigma_{TWB}(\omega)& =\frac{M_{TWB}(\omega)}{W_{TWB}}
\end{align}
where $M_{TWB}(\omega)$ and $V_{TWB}(\omega)$ denote the bending moment and shear force at the tower base; $\mathbf{C}^{TOT}(7,7)$ is the total stiffness associated with the tower-top deflection DoF, $q_7(\omega)$; $h_{TT}$ is the height to the tower top, i.e., the deformable length of the tower; while $W_{TWB}$ is the section modulus of the tower base. Additional details on the dynamic model can be found in \cite{Ferri2022,Ferri2023}. 


\subsection{Metocean Data Analysis} 
\begin{figure}[!htb]
\unskip
\centering
\includegraphics[width = 0.44\textwidth]{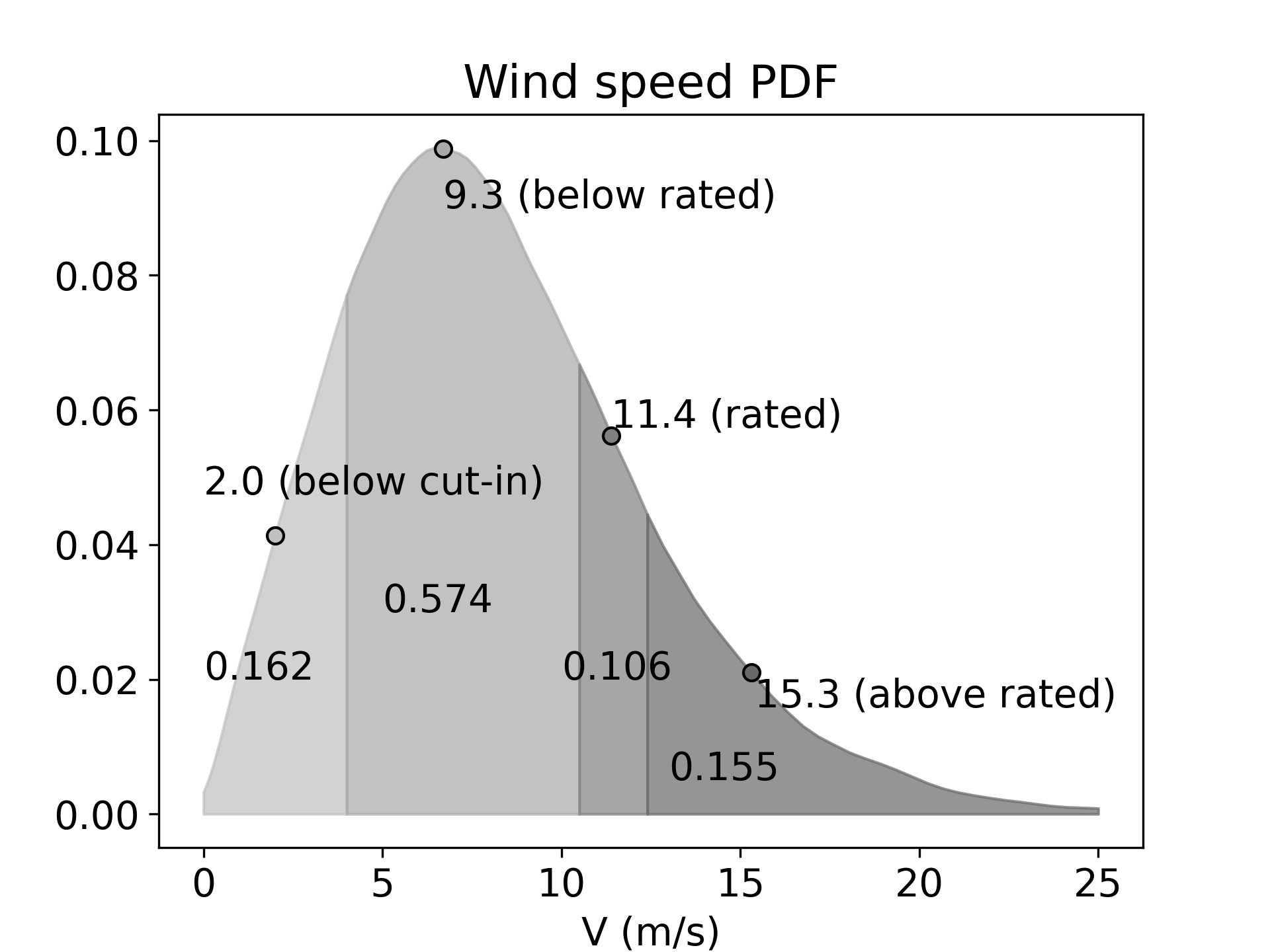}	
\caption{Definition of four mean wind speed bins. \label{fig:wind}}
\end{figure}

Hourly metocean data on mean wind speed ($U$) at \SI{10}{\meter} above MSL, significant wave height ($H_s$) and wave peak period ($T_p$) from January 1993 to June 2021 were collected at the selected site in the Mediterranean Sea, near the Sicilian coast. A power-law vertical profile is assumed to convert $U$ to $V$,  the mean wind speed at the hub height, \SI{90}{\meter} above the mean water level. Based on characteristics of the NREL 5-MW wind turbine control system, we study the system response and fatigue damage in different operating states; accordingly, we consider four $V$ bins: below cut-in ($V$ $<$ \SI{3}{\meter/\second}), below rated (\SI{3}{\meter/\second} $\leq V$ $<$ \SI{10.5}{\meter/\second}), near rated (\SI{10.5}{\meter/\second} $\leq V$ $<$ \SI{12.4}{\meter/\second}) and above rated ($V \geq$ \SI{12.4}{\meter/\second}). Additionally, a representative mean wind speed for each bin is selected as is indicated in Fig.~\ref{fig:wind}.

Wave height and period data for each wind speed bin are considered using kernel density functions to describe their joint distributions~\cite{scott2015multivariate}: 
\begin{equation}
    \hat{p}(H_s,T_p) = \frac{1}{nh_{1}h_{2}}\sum_{i=1}^{n}\left(Ker\left(\frac{H_s-H_{s,i}}{h_{1}}\right)\cdot Ker\left(\frac{T_p-T_{p,i}}{h_{2}}\right)\right)
\label{eq:kde}
\end{equation}
where $n$ represents the number of data points $(H_{s,i},T_{p,i})$ in the wind speed bin, $Ker$ refers to the selected non-negative Gaussian kernel function, and $h_{1},h_{2}>0$ are smoothing parameters that define the bandwidth. 

\begin{figure}[!htb]
\centering
\unskip
\includegraphics[width = \textwidth]{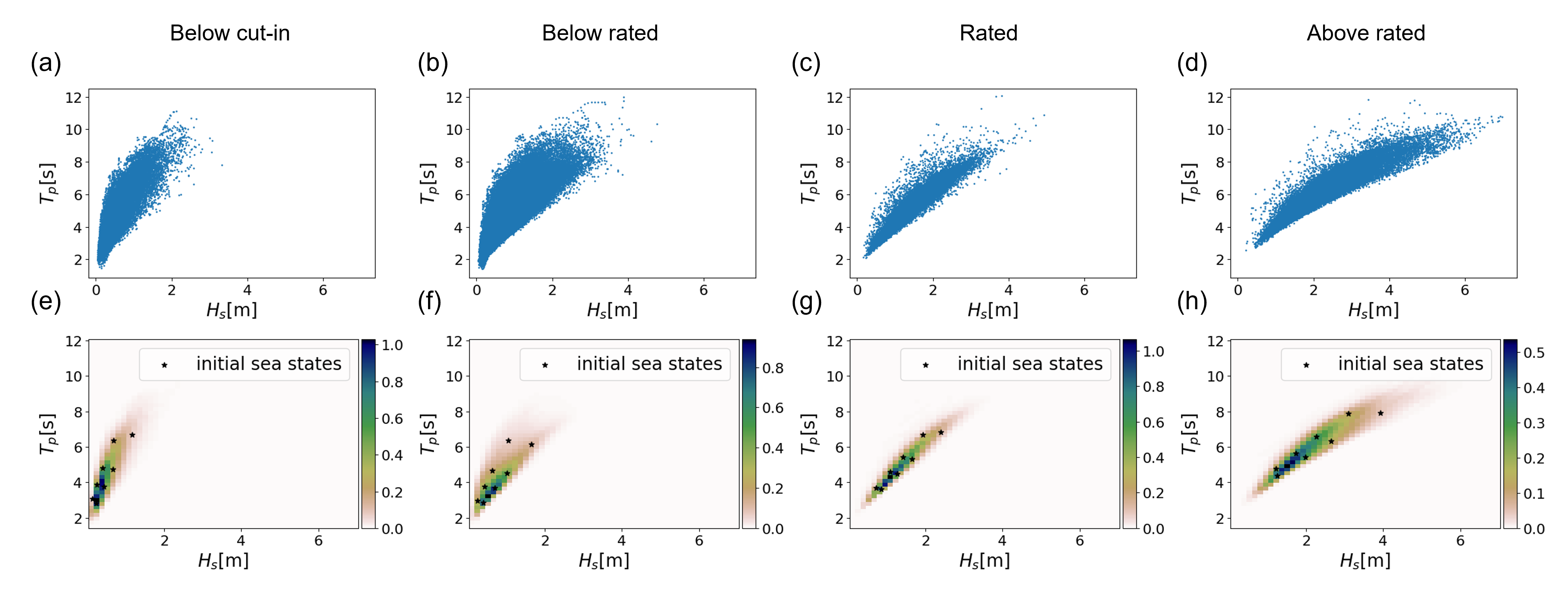}
\caption{(a-d) Wave height and peak period data; and (e-h) kernel density function fits for sea states in the 4 wind speed bins.\label{fig:wave}}
\end{figure}

Using the joint distribution of $(H_{s},T_{p})$ in each wind bin, we select 8 representative sea states that serve as the initial training data sets for the GPR model. Specifically, two principal components derived from the bivariate data are identified to efficiently cover the entire $(H_{s},T_{p})$ domain of the sea states. The first principal component (with the greatest variance) is divided into four intervals and the second principal component is divided into two intervals with equal probability based on the marginal component distribution. Representative $(H_s,T_p)$ pairs are selected from these 4$\times$2 grids based on computed joint density-weighted centers. Figure~\ref{fig:wave} shows the wave height and peak period data points and kernel density-based representative sea states for each wind speed bin.  Such kernel density functions fit to metocean data allow non-parametric formulations for many applications when dealing with fatigue as well as ultimate limit states~\cite{manuel2018alternative}.

\subsection{Fatigue Damage Assessment} 
For any variable-amplitude stress process with ranges, $S$, defined by a probability density function (PDF), $p(s)$, the short-term fatigue damage experienced over exposure time, $T$, can be computed as follows:
\begin{equation}
\label{eq:damage} 
D(T)=n(T) \int_{0}^{\infty}\frac{1}{N(S)}p(S)dS 
\end{equation}
where $n(T)$ is the total number of cycles encountered over $T$; $N(S)$ represents the number of cycles of stress amplitude, $S$, that are needed for failure.  For a single-slope S-N curve with parameters, $K_a$ and $b$, we have $N(s)=K_a \, S^{-b}$.  Now, given the one-sided power spectral density function, $G(f)$, of the stress process, one can readily compute spectral moments, $m_j$, as follows:
\begin{equation} 
\label{eq:moment}  
m_j=\int_{0}^{\infty} f^j \,G(f) \,df \  
\end{equation}

Dirlik's method~\cite{dirlik1985application,Dirlik2021Metals,Ragan2007WindEngineering} allows simplified fatigue damage estimation in which $p(S)$ is computed using the spectral moments, $m_j$.  Thus, we have:
\begin{equation}
\label{eq:Dirlik} 
p(s)=\frac{1}{2\sqrt{m_0}}
\left[
\frac{G_1}{Q}e^{-\frac{Z}{Q}}+\frac{G_2Z}{R^2}e^{-\frac{z^2}{2R^2}}+G_3Ze^{-\frac{Z^2}{2}}
\right]
\end{equation}
where $Z=S/2\sqrt{m_0}$, 
$G_1 = 2\left(x_m-\alpha_2^2\right)/\left(1+\alpha_2^2\right)$, 
$G_2 = \left({1-\alpha_2-G_1+G_1^2}\right)/\left({1-R}\right)$ and    
$G_3 = 1-G_1-G_2$, with $x_m=(m_1/m_0)\sqrt{(m_2/m_4)}$,  $\alpha_2=m_2/\sqrt{m_0m_4}$, 
$Q={1.25(\alpha_2-G_3-G_2R)}/{G_1}$
and 
$R = \left(\alpha_2-x_m-G_1^2\right)
\left(
{1-\alpha_2-G_1+G_1^2}
\right)$.

Finally, using this stress amplitude PDF $p(s)$, the fatigue damage, $D(T)$ for an exposure time, $T$, can be estimated as follows \cite{Dirlik2021Metals}:
\begin{equation}
\begin{split}
&D(T)=\frac{\nu_pT}{K_a}\left(\sqrt{m_0}\right)^b
\left[
G_1Q^b\Gamma(1+b)+
(\sqrt{2})^b\Gamma\left(1+\frac{b}{2}\right)+
\left(G_2|R|^b+G_3\right)
\right],\\ \; &\text{where} \; \nu_p=\sqrt{m_4/m_2}  
\end{split}
\end{equation}

To make it convenient to compare fatigue damage across different sea states for the same exposure time, $T$(sec), we introduce the concept of a constant-amplitude ``$F$-Hz Damage-Equivalent Load'' that, by definition, leads to the same damage over time, $T$, as occurs with the variable-amplitude stress cycles in that same exposure time.  If we select $F=\SI{1}{\hertz}$, we can then compute short-term 1-Hz damage-equivalent load ($DEL_{\text{1-Hz}}$) estimates for any sea state because
$D(T) = (1 \cdot T) / N(DEL_{\text{1-Hz}}) = 
(1 \cdot T)/K_a \cdot DEL_{\text{1-Hz}}^{b}$, and thus we have, for $T$ in seconds: 
\begin{equation}
DEL_{\text{1-Hz}}= 
\sqrt{m_0} \cdot \nu_p^{-b}
\left[
G_1Q^b\Gamma(1+b)+
(\sqrt{2})^b\Gamma\left(1+\frac{b}{2}\right)+
\left(G_2|R|^b+G_3\right)
\right]^{-b}
\end{equation}
In this study, GPR response surfaces are constructed and iteratively refined using $DEL_\text{1-Hz}$ from intermediate training sets. These response surfaces allow predictions of $DEL_\text{1-Hz}$ for each sea state. For long-term damage estimation, the surfaces should accumulate contributions to short-term fatigue damage from the different sea states. The predicted value of the short-term damage $\hat{D}$ can be written as a function of time $T$ and the variables, $V$, $H_s$ and $T_p$. Finally, the long-term fatigue damage, $LTD(T)$, over an exposure period, $T$, can be estimated as follows:
\begin{equation}
LTD(T)=\int \hat{D}(T,V,H_s,T_p)\hat{p}(V,H_s,T_p)d(V,H_s,T_p)
\end{equation}
where $\hat{p}(V,H_s,T_p)$ represents the joint PDF at each sea state, which is in turn estimated by multiplying the occurrence probability of each wind bin with $\hat{p}(H_s,T_p)$.

\subsection{Gaussian Process Regression} 
We have seen that in order to evaluate the long-term fatigue damage, we need to accumulate $DEL_\text{1-Hz}$ estimates for the real metocean environment at the installation site over the expected or planned service life of the FOWT. This requires a very large number of simulations. We seek to accelerate the computations using Gaussian process regression \cite{williams1995gaussian,RobertsPTSA}; a model thus built using GPR can then serve as a surrogate for exhaustive coupled frequency-domain model evaluations and subsequent fatigue assessment. The surrogate must be trained to ``learn'' how to relate input sea state variables $(H_s,T_p)$ to the output damage, $DEL_\text{1-Hz}$, for each wind speed bin, using simulations results available only for a small subset of representative sea states. 

Suppose $q$ sea states are selected as training data, $X=\{\mathbf{x}_1,\ldots,\mathbf{x}_q\}$ where $\mathbf{x}_i=(H_{s_i},T_{p_i})$. From the preceding discussion, we calculate the corresponding $DEL_\text{1-Hz}$ estimates, $\mathbf{y}=\{y_1,\ldots,y_q\}$ where $y_i$ is $DEL_\text{1-Hz}$ of $\mathbf{x}_i$. Then the possibly complex and nonlinear relationship between input $X$ and output $\mathbf{y}$ can be approximated in terms of a function $f$ with Gaussian noise $\epsilon \sim N(0,\sigma^2)$, i.e. $y_i = f(\mathbf{x}_i) + \epsilon$. We model this function using the Gaussian process prior, $f(\mathbf{x}) \sim GP(\mu(\mathbf{x}),k(\mathbf{x},\mathbf{x}'))$. The prior mean function is modeled as a (non-zero) constant, $\mu(\mathbf{x}) = C$, and the prior covariance on two samples $\mathbf{x},\mathbf{x}'$ is modeled by the popular radial basis function kernel, $k(\mathbf{x},\mathbf{x}')=s\cdot\exp{\left(-(||\mathbf{x}-\mathbf{x}'||^2)/{2l^2}\right)}$. First, we must learn the hyperparameters $\theta = (\sigma,s,l)$ from the training set, $(X,\mathbf{y})$, by minimizing the negative log marginal likelihood:
\begin{equation}
\begin{split}
&L(\theta|X,\mathbf{y})\propto\log{|\hat{\mathbf{K}}_{XX}|}-\mathbf{y}^{\top}\hat{\mathbf{K}}_{XX}^{-1}\mathbf{y},\\
&\frac{dL}{d\theta} = \mathbf{y}^{\top}\hat{\mathbf{K}}_{XX}^{-1}\frac{d\hat{\mathbf{K}}_{XX}}{d\theta}\hat{\mathbf{K}}_{XX}^{-1}\mathbf{y}+\text{tr}\left(\hat{\mathbf{K}}_{XX}^{-1}\frac{d\hat{\mathbf{K}}_{XX}}{d\theta}\right)
\end{split}
\end{equation}
where $\mathbf{K}_{XX}$ denotes the matrix containing all pairs of the kernel entries---i.e.,~$[K_{XX}]_{ij}=k(\mathbf{x}_i,\mathbf{x}_j)$. The hat symbol denotes an added diagonal: $\hat{\mathbf{K}}_{XX}=\mathbf{K}_{XX}+\sigma^2 I$. Then, predictions can be made by the predictive posterior distribution, $p(f(\mathbf{x}^{*})|X,\mathbf{y})$. Its predictive mean and covariance are given as:
\begin{equation}
\begin{split}
&\mu_{f|X,\mathbf{y}}(\mathbf{x}^{*})=\mu(\mathbf{x^{*}}) + \mathbf{k}_{X\mathbf{x}^*}^\top\hat{\mathbf{K}}_{XX}^{-1}\mathbf{y}\\
&k_{f|X,\mathbf{y}}(\mathbf{x}^*,\mathbf{x}^*{}') = k_{\mathbf{x}^*\mathbf{x}^*{}'} - \mathbf{k}_{X\mathbf{x}^*}^\top\hat{\mathbf{K}}_{XX}^{-1}\mathbf{k}_{X\mathbf{x}^*{}'}
\end{split}
\end{equation}
where $\mathbf{k}_{X\mathbf{x}^*}$ denotes kernel values using the training input $X$ and a test point $\mathbf{x}^*$---i.e.,~$[\mathbf{k}_{X\mathbf{x}^*}]_i = k(\mathbf{x}_i,\mathbf{x}^*)$ \cite{gardner2018gpytorch}. Using the predictive posterior, $DEL_\text{1-Hz}$, for any test point, $\mathbf{x}^*$ (i.e., for any sea state), can be predicted. In this way, long-term fatigue damage can be easily accumulated using historical records or randomly sampled sea state data from a joint distribution model based on metocean statistics. 

With such an approach, the accuracy of the long-term fatigue damage evaluation depends on the accuracy of the regression.  Note that  $k_{f|X,\mathbf{y}}(\mathbf{x}^*,\mathbf{x}^*)$ indicates the variance of the predicted value at $\mathbf{x}^*$. The predictive value of $DEL_\text{1-Hz}$ for $\mathbf{x}^*$ is expected to follow a normal distribution, $N(\mu (\mathbf{x}^*),\sigma^2(\mathbf{x}^*))$. New training data points can be selected based on the predictive variance \cite{pasolli2011gaussian} and this predictive variance would decrease over iterations. Because the joint probability of occurrence and the importance (to damage) are in general different for distinct sea states, a variance-based weighted learning algorithm should be adopted \cite{yue2020active}. In this study, we consider the confidence interval of weighted short-term damage as an indicator of uncertainty. Given a $z$ score, $\gamma$, the confidence interval of the predicted $LTD(T)$ can be estimated as:

\begin{equation}
\begin{split}
&\frac{T}{K_a}\int [\mu(\mathbf{x}^*)-\gamma\sigma(\mathbf{x}^*)]^b\hat{p}(\mathbf{x}^*)d\mathbf{x}^* <\frac{T}{K_a}\int \mu(\mathbf{x}^*)^b\hat{p}(\mathbf{x}^*)d\mathbf{x}^*\\
&<\frac{T}{K_a}\int [\mu(\mathbf{x}^*)+\gamma\sigma(\mathbf{x}^*)]^b\hat{p}(\mathbf{x}^*)d\mathbf{x}^*
\end{split}
\end{equation}
where $\hat p(\mathbf{x}^*)$ is the PDF evaluated at the sea state $\mathbf{x}^*$ . In this study, $b$ is assumed to be 3; thus, we define the width of the confidence interval for short-term damage ,$CI(T,\mathbf{x}^*)$, as the predictive uncertainty at each sea state:
\begin{equation}
CI(T,\mathbf{x}^*) = \frac{T}{K_a} [6\mu(\mathbf{x}^*)^2\gamma\sigma(\mathbf{x}^*)+2\gamma^3\sigma(\mathbf{x}^*)^3]\hat{p}(\mathbf{x}^*)d\mathbf{x}^*
\end{equation}

In each iteration, a new sea state, $\mathbf{x}_{q+1}$, with the highest $CI(T=3600\text{s})$ is added to the training set $X$ for the next iteration:

\begin{equation}
\mathbf{x}_{q+1}=\underset{\mathbf{x}^*}{\mathrm{argmax}}(CI(T=3600\text{s},\mathbf{x}^*))
\end{equation}

\begin{figure}[!htb]
\unskip
\centering
\includegraphics[width = 1\textwidth]{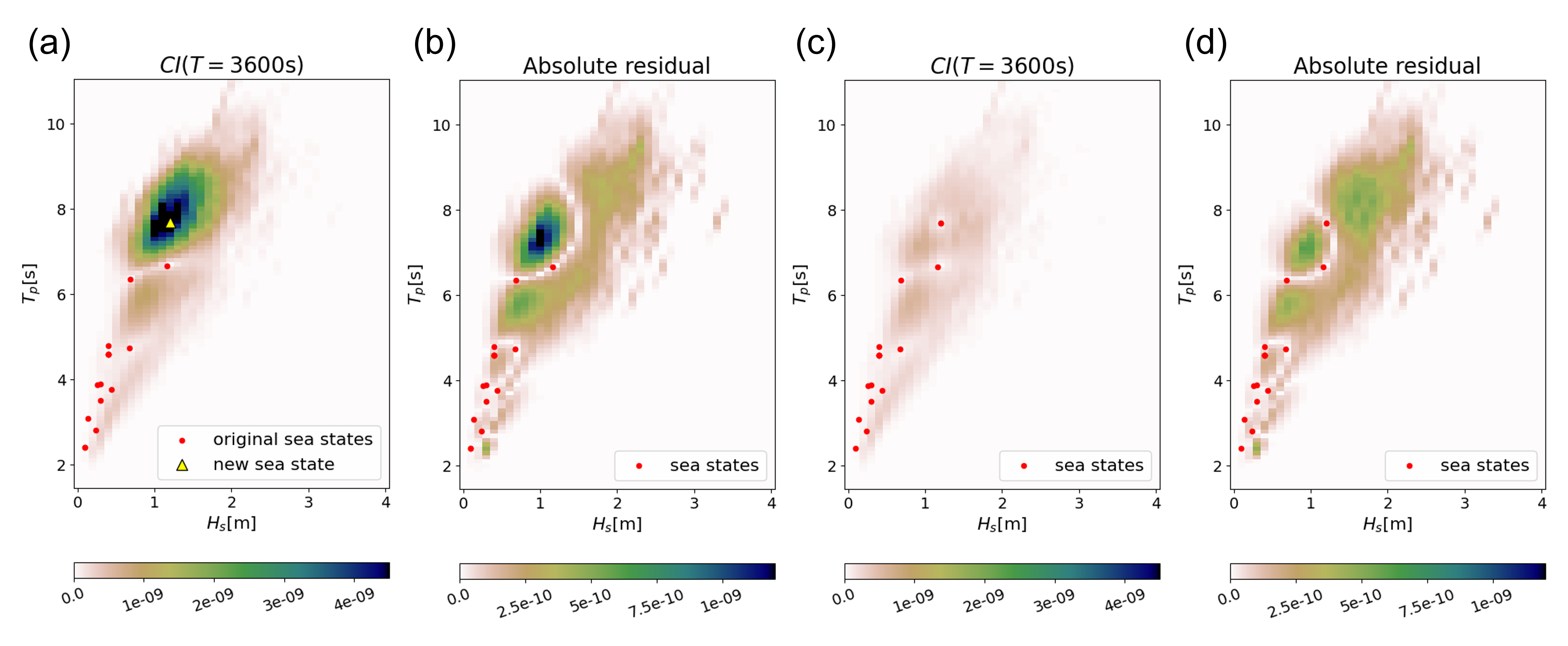}
\caption{An illustration of GPR surface refinement from (a-b) $i^{th}$ iteration to (c-d) $(i+1)^{th}$ iteration. \label{fig:iteration}}
\end{figure}

\begin{figure}[!htb]
\unskip
\centering
\includegraphics[width = 1\textwidth]{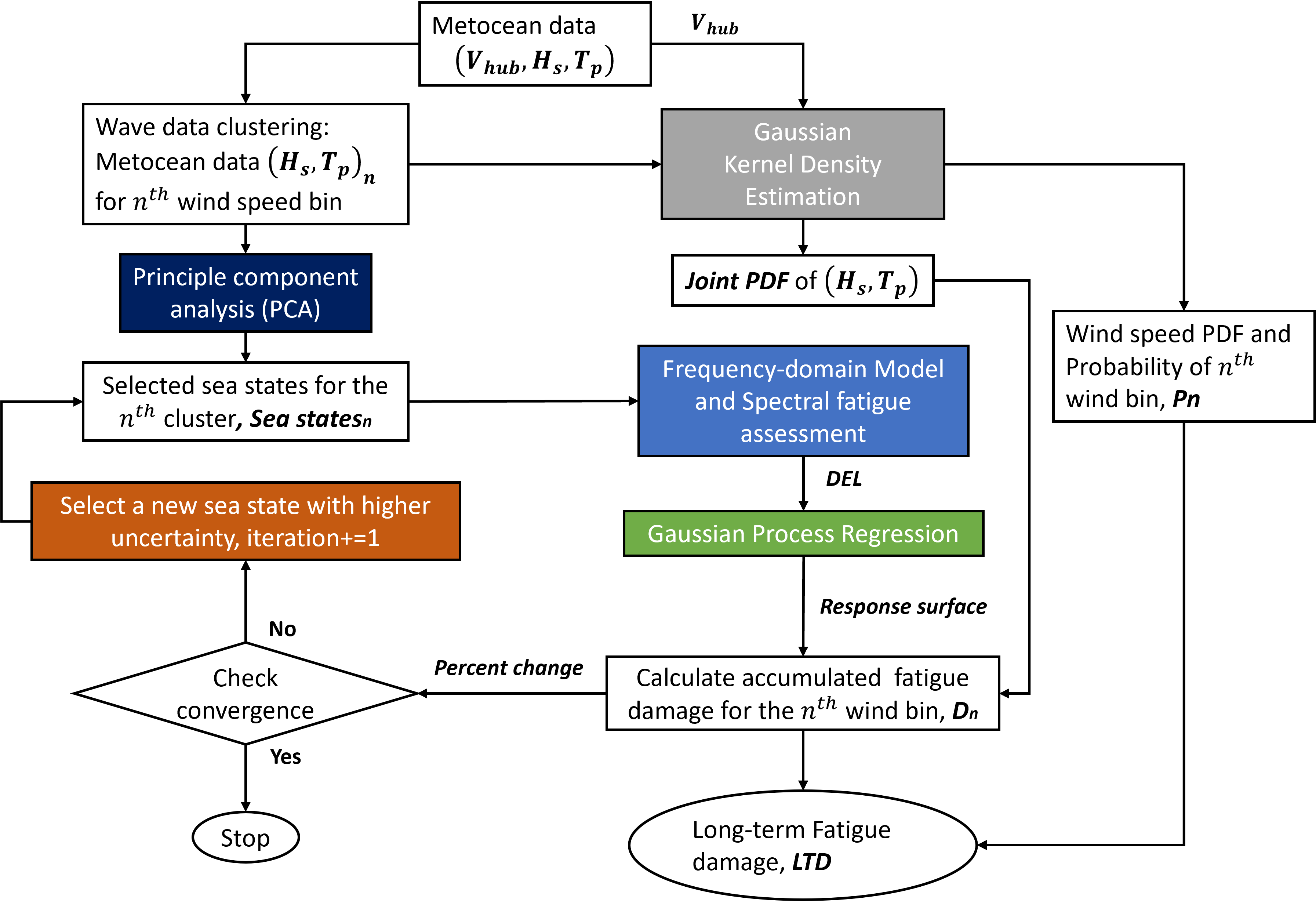}
\caption{Framework for efficient long-term fatigue damage assessment based on a Gaussian process regression surrogate model. \label{fig:framework}}
\end{figure}

Figure~\ref{fig:iteration} illustrates how the new sea state is selected to reduce the absolute residual $|\hat{D}(T,\mathbf{x}^*)\hat{p}(\mathbf{x}^*)-D(T,\mathbf{x}^*)\hat{p}(\mathbf{x}^*)|$. In Fig.~\ref{fig:iteration}(a), the sea state with the highest $CI(T=\SI{3600}{\second})$ in the $i^{th}$ iteration is added to the training data for the next iteration (Fig.~\ref{fig:iteration}(c)), thus reducing the residual significantly, as shown in Figs.~\ref{fig:iteration}(b) and (d). The proposed framework for efficient long-term fatigue damage assessment based on a surrogate Gaussian process regression model is presented in Fig.~\ref{fig:framework}. The metocean data analysis defines the initial sea states used to train the GPR model. The GPR model for $DEL_\text{1-Hz}$ of each wind speed bin is updated in every iteration with an additional sea state. When the response surface for each wind bin has converged, overall long-term fatigue damage can be calculated.


\section{Results and Discussion}
\subsection{Coupled Simulations in the Frequency Domain and Fatigue Damage Assessment}
Simulations in the frequency domain are performed applying the standard wind and wave power spectra as described in Fig.~\ref{fig:FD}. The results obtained include the amplitude of motions of the platform and the tower-top deflection; from these, power spectra for other relevant response processes are computed. For instance, the assumption of rigid-platform motions allows estimation of the PSD of a mooring line's tension at the fairlead from the PSDs for surge, heave and pitch, following the procedure described in \cite{Ferri2022}. Similarly, the PSD of the bending moment at the tower base is calculated from the PSD of the tower top deflection~\cite{Ferri2023}. Then, using the cable cross-sectional area and the tower base cross-sectional dimensions, stress PSDs are obtained.

For fatigue damage assessment, the parameter S-N curve parameter, $K_a$ (in MPa$^3$), is assumed to be $1.46\times 10^{12}$ and $1.20\times10^{11}$ for the tower base and mooring line, respectively \cite{dnv2016dnvgl}. Extensive full-grid simulations (over the range of wave heights and periods) that cover all sea states are conducted to obtain exact fatigue damage and to serve as the reference solution, against which the surrogate model is assessed. 
This reference 1-hour long-term damage $LTD(T=\SI{3600}{\second})$ is $2.997\times10^{-5}$ for the tower base and $4.118\times10^{-5}$ for the mooring line. 

\begin{figure}[!htb]
\unskip
\centering
\includegraphics[width = \textwidth]{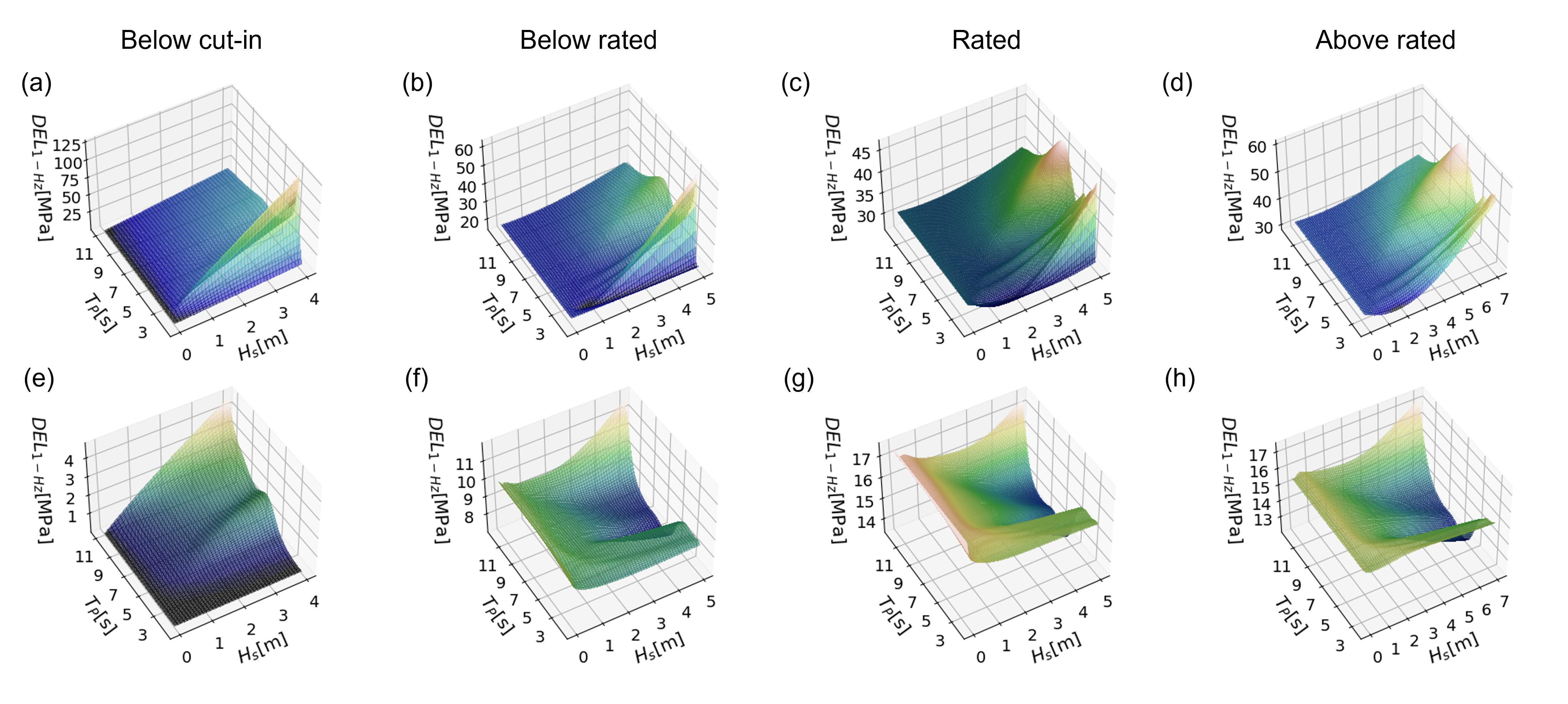}
\caption{Exact $DEL_\text{1-Hz}$ from full-grid simulation for (a-d) the tower base and (e-h) the mooring line.\label{fig:exact}}
\end{figure}

Figure~\ref{fig:exact} summarizes $DEL_\text{1-Hz}$ variations with $H_s$ and $T_p$ for all the wind speed bins that result from the simulations for the tower base and mooring line loads.  
For the tower base, higher fatigue loads are induced at higher wind speeds and higher wave heights. This is associated with the greater gyroscopic damping of the rotating rotor around the rated wind speed, which has the effect of mitigating rotational motions of the entire system~\cite{Ferri2022}. 
In contrast, the greatest fatigue damage in the mooring line is found to occur around the rated wind speed. Under such conditions, larger surge motions are reached; consequently, the moorings become tauter and stiffer, leading to an increase in the tension in the cables. Indeed, the largest effects related to this phenomenon are seen for the rated wind speed case (Fig.~\ref{fig:exact}(g)), i.e., when the thrust force is maximum. 
Catenary line dynamics are dominated by the system response in the platform motion DoFs~\cite{Ferri2022}; fatigue damage tends to increase for longer spectral periods where the wave spectrum contains more energy and frequency bands close to the platform eigenfrequencies then are amplified in the response. Moreover, the sharp increase in $DEL_\text{1-Hz}$ values seen for $H_s$ values near zero in Figs.~\ref{fig:exact}(f-h) is probably related to reduction of the hydrodynamic viscous drag damping because these drag forces are positively correlated with wave and motion amplitudes. 

The nonlinearity in $DEL_\text{1-Hz}$ values across different sea states is evident in the irregular response surfaces; the complexity of wind and wave interaction is also clear. 
Focusing on Figs.~\ref{fig:exact}(a-d), we see sharp peaks for sea states with  spectral periods varying from \SI{2}{\second} to \SI{4}{\second} that are related to the dynamic amplification occurring at those periods; the peak around \SI{2.25}{\second} is mostly caused by resonance with the first tower eigenfrequency, and it tends to be prominent for below-rated wind conditions (Figs.~\ref{fig:exact}(a-b)) due to lower rotor gyroscopic damping contribution. 

\begin{figure}[!htb]
\unskip
\centering
\includegraphics[width = \textwidth]{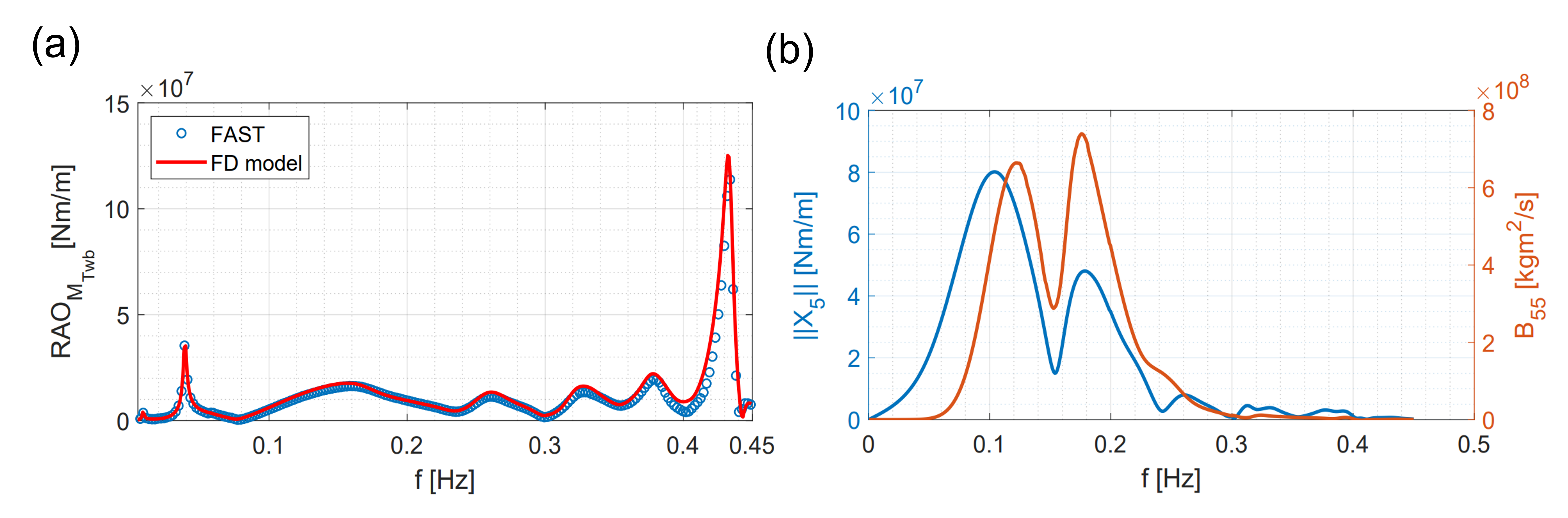}
\caption{(a) Tower-base bending moment response amplitude operator in the parked state; and (b) hydrodynamic force amplitude and radiation damping in pitch.}\label{fig:hp}
\end{figure}

In order to further investigate system response peaks at different frequencies, Fig.~\ref{fig:hp}(a) shows plots of the response amplitude operator, i.e., the linearized transfer function of the coupled system \cite{Ferri2022}, for the tower-base bending moment in the parked state. blue circles refer to results obtained from time-domain simulations in FAST, while the solid red line applies to results obtained with the FD model used in this work. 
Dynamic amplification with a very similar pattern as in Figs.~\ref{fig:exact} (a-d) is clearly seen over the range of periods \SI{2.5}{\second}-\SI{4}{\second} (i.e., \SI{0.25}{\hertz}-\SI{0.4}{\hertz}). 
Such behavior is related to the floater properties, reflecting coupling between the tower and platform and the influence of the hydrodynamic force ($\textbf{X}(\omega)$ in Eq.~\eqref{eq:eq1}) amplitude in the pitch direction (see the blue line in Fig.~\ref{fig:hp}(b)). Moreover, the pitch radiation damping ($\textbf{B}(\omega)$ in Eq.~\eqref{eq:eq1}) of the platform tends to decrease and vanishes for frequencies above \SI{0.25}{\hertz} (see the orange line in Fig.~\ref{fig:hp}(b)), leading to an increase in the dynamic response of the system. The nonlinearity evident in the short-term response surfaces for fatigue damage can be explained by the structural system's dynamic characteristics; additional information about the structural response could be included in training surrogate models for similar fatigue damage studies in future work.

\subsection{Gaussian Process Regression}
\begin{figure}[!htb]
\unskip
\centering
\includegraphics[width = 1\textwidth]{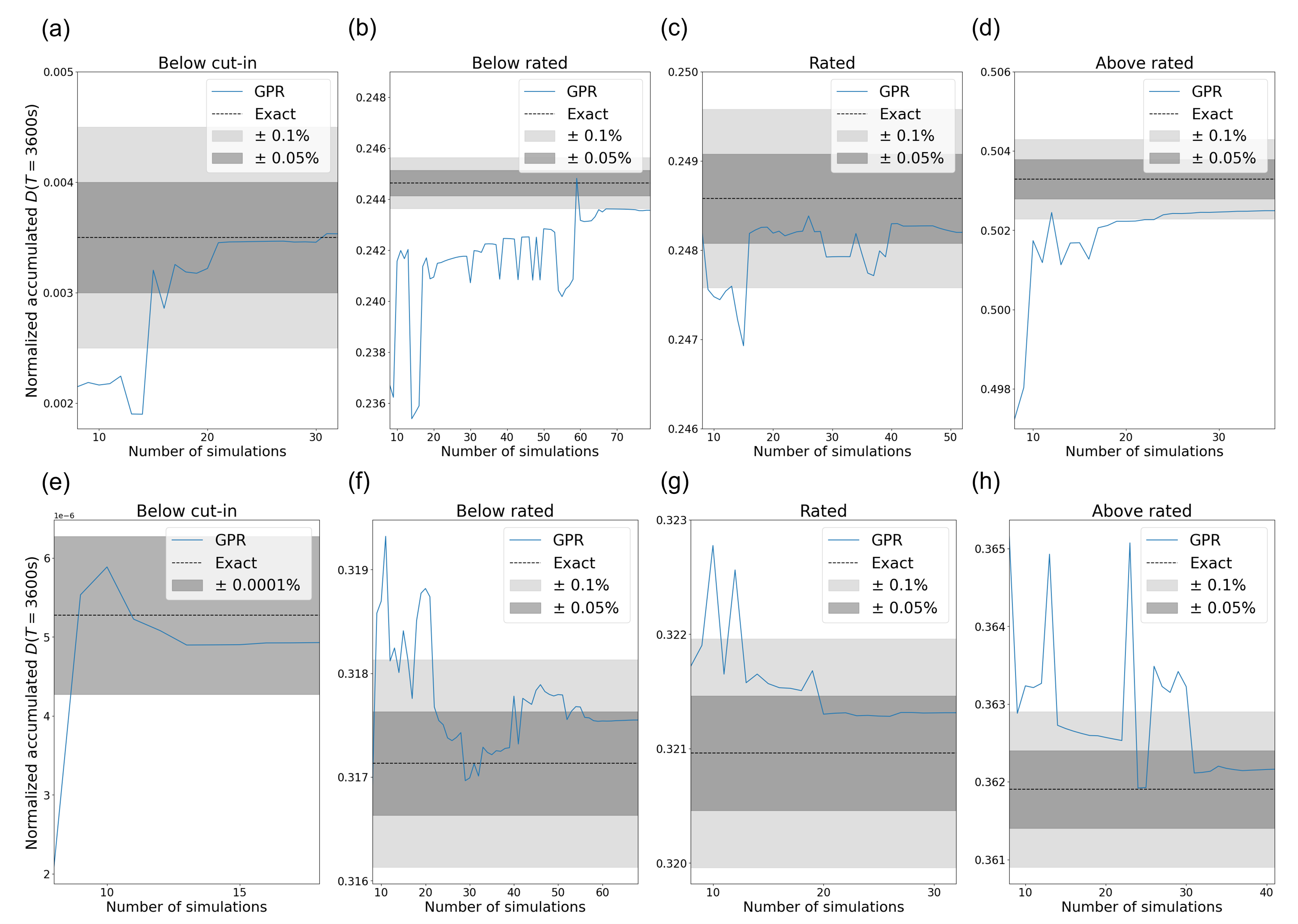}
\caption{Convergence of normalized accumulated damage in each wind bin for (a-d) the tower base and (e-f) the mooring line.\label{fig:AD}}
\end{figure}

Initially, eight sea states are selected, thus making up the training data set for each of the four wind speed bins.  This implies that only 32 FD simulations are needed in the first iteration. More simulations are then added to the original training data set so that the solution is gradually improved. When the change in accumulated short-term damage for each wind speed bin falls below 0.01\% of $LTD(T=\SI{3600}{\second})$ over 10 successive iterations, the damage estimate is assumed to have converged. This strategy ensures that new simulations only focus on the wind speed bin that contributes most to overall fatigue damage. Figure~\ref{fig:AD} shows the pattern of convergence of accumulated damage for each wind speed bin. The accumulated damage is normalized with respect to $LTD(T=\SI{3600}{\second})$ in order to compare the relative importance across the wind speed bins. For example, the accumulated damage for the below cut-in wind speed bin, when considering the tower-base bending moment, is less than $5\times 10^{-6}$ of total $LTD(T=\SI{3600}{\second})$; therefore, fewer simulations are conducted in this wind bin. The final result with each surface has errors that fall to merely 0.05-0.1\% of the exact $LTD(T=\SI{3600}{\second})$. 

\begin{figure}[!htb]
\unskip
\centering
\includegraphics[width =1\textwidth]{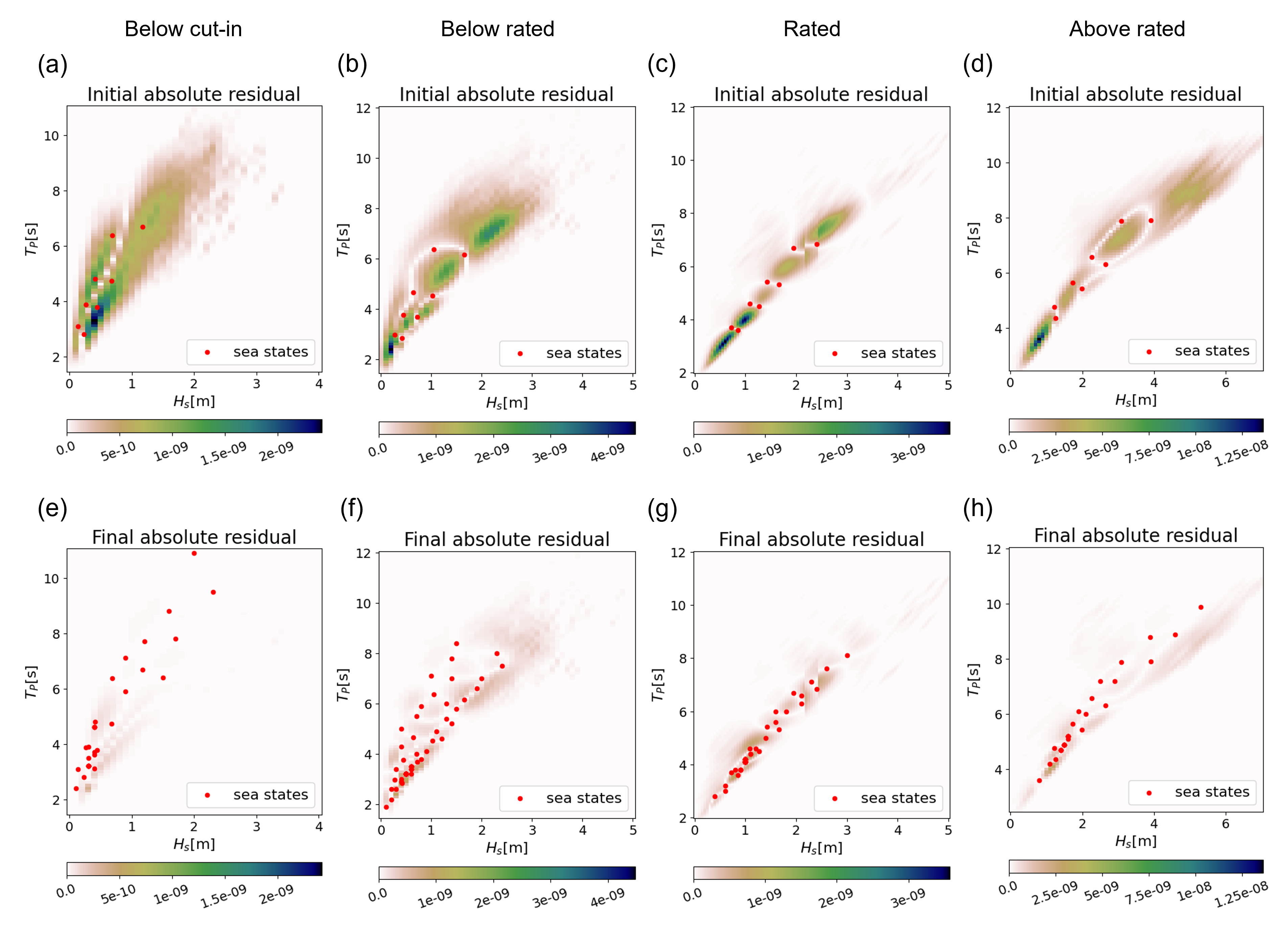}
\caption{Comparison of (a-d) initial absolute residual and (e-h) final absolute residual after GPR response surfaces refinement for short-term fatigue damage at the tower base. \label{fig:residual1}}
\end{figure}

\begin{figure}[!htb]
\unskip
\centering
\includegraphics[width = 1\textwidth]{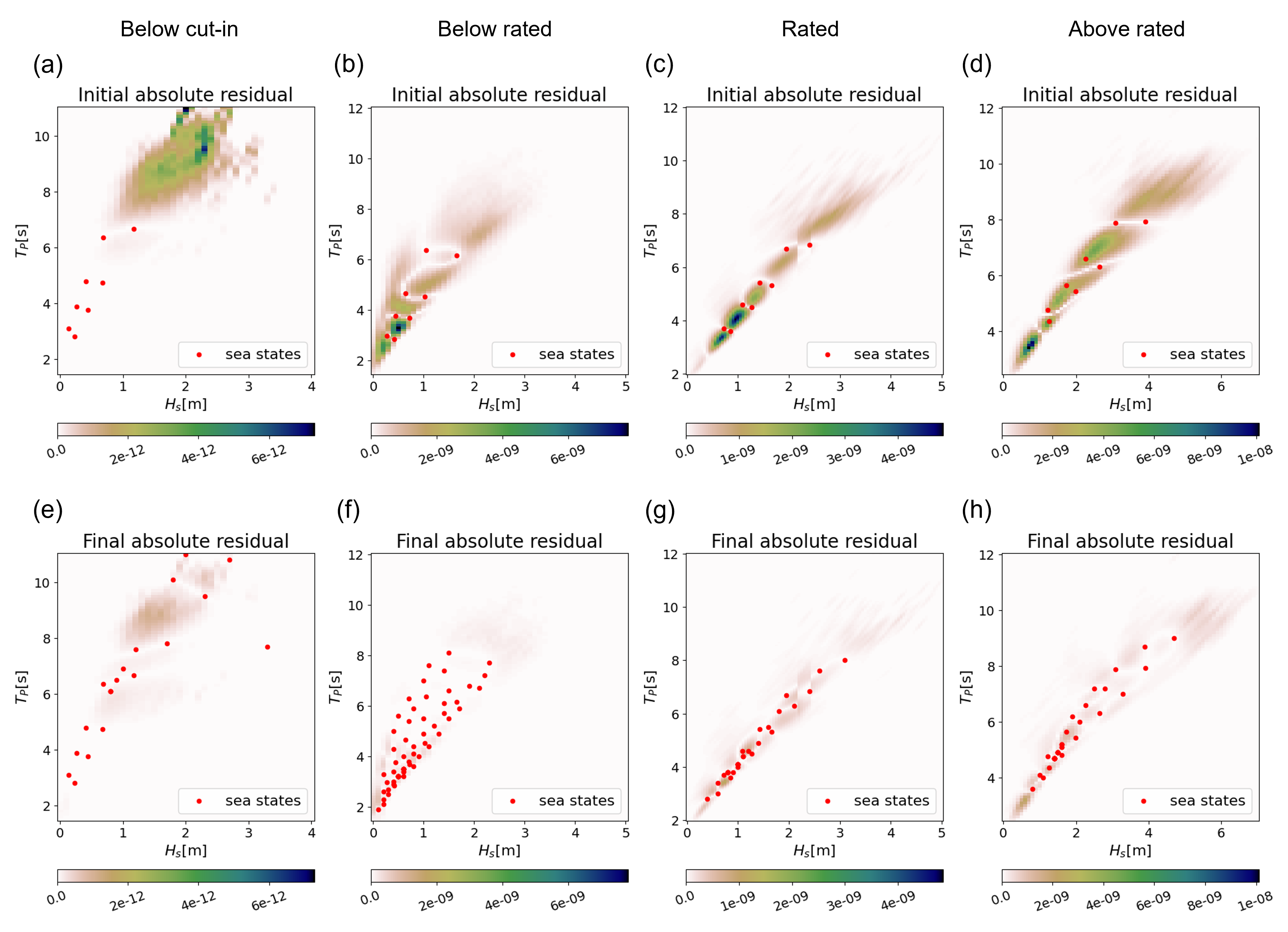}
\caption{Comparison of (a-d) initial absolute residual and (e-h) final absolute residual after GPR response surfaces refinement for short-term fatigue damage in the mooring line at the fairlead. \label{fig:residual2}}
\end{figure}

\begin{figure}[!htb]
\unskip
\centering
\includegraphics[width = \textwidth]{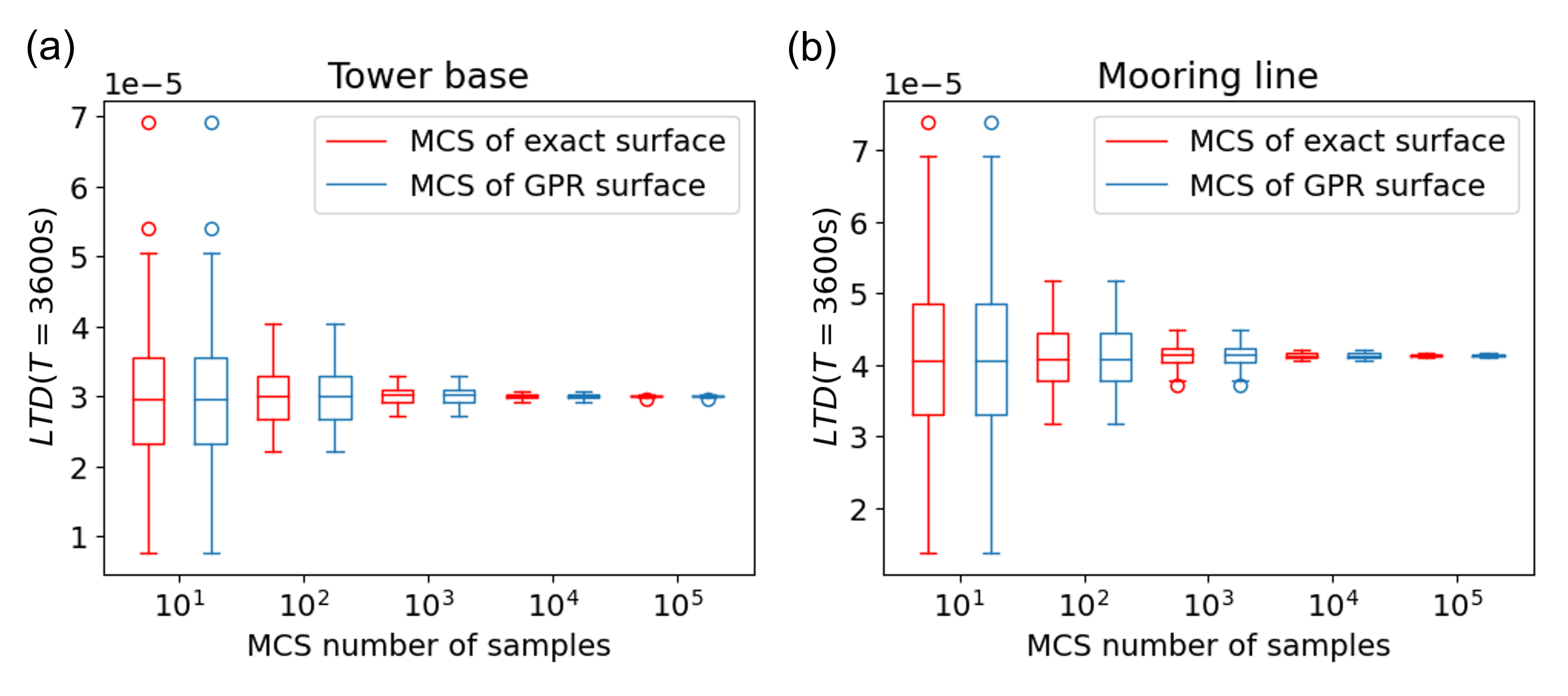}
\caption{Box plots for $LTD(T=\SI{3600}{\second})$ estimates with different numbers of MCS simulations that employ the GPR surfaces and the exact surfaces for (a) the tower base bending moment; and (b) the mooring line fairlead tension.\label{long-term}}
\end{figure}

Figures~\ref{fig:residual1} and \ref{fig:residual2} compare absolute residual short-term fatigue damage estimates between the initial surfaces and the final converged surfaces in each wind bin. As can be confirmed visually in both figures, the GPR response surfaces for $\hat{D}\times \hat{p}$  are very close to the exact solution derived from extensive full-grid simulations. \textcolor{black}{For the tower base, the maximum absolute residual of each surface ranged from $2.4\times 10^{-9}$ to $1.4\times 10^{-8}$ initially, and then converged to the final surfaces with maximum absolute residual ranging from $3.6\times 10^{-10}$ to $3.2\times 10^{-9}$.  The maximum residual for each surface is reduced by at least 75\% through the refinement process. Similar results are found for converged surfaces for the mooring lines. } 
Moreover, by carrying out Monte Carlo simulations (MCS) that employ both surfaces separately, one can again confirm their similarity. In Fig.~\ref{long-term}, MCS with these two surfaces, using different numbers of samples (ranging from $10^1$ to $10^5$), are repeated 100 times to assess variability in predictions, indicated by means of box plots as well as convergence rates. These results show the excellent performance of the GPR surfaces in predicting the exact solution. Since the box plots with the GPR surfaces are almost the same as those for the costlier exact surfaces, the similarity can be confirmed. As the GPR surrogate model is generated from fewer simulations, it is considered more efficient compared to the use of exact surfaces in the estimation of long-term fatigue damage.

\begin{figure}[!htb]
\unskip
\centering
\includegraphics[width = \textwidth]{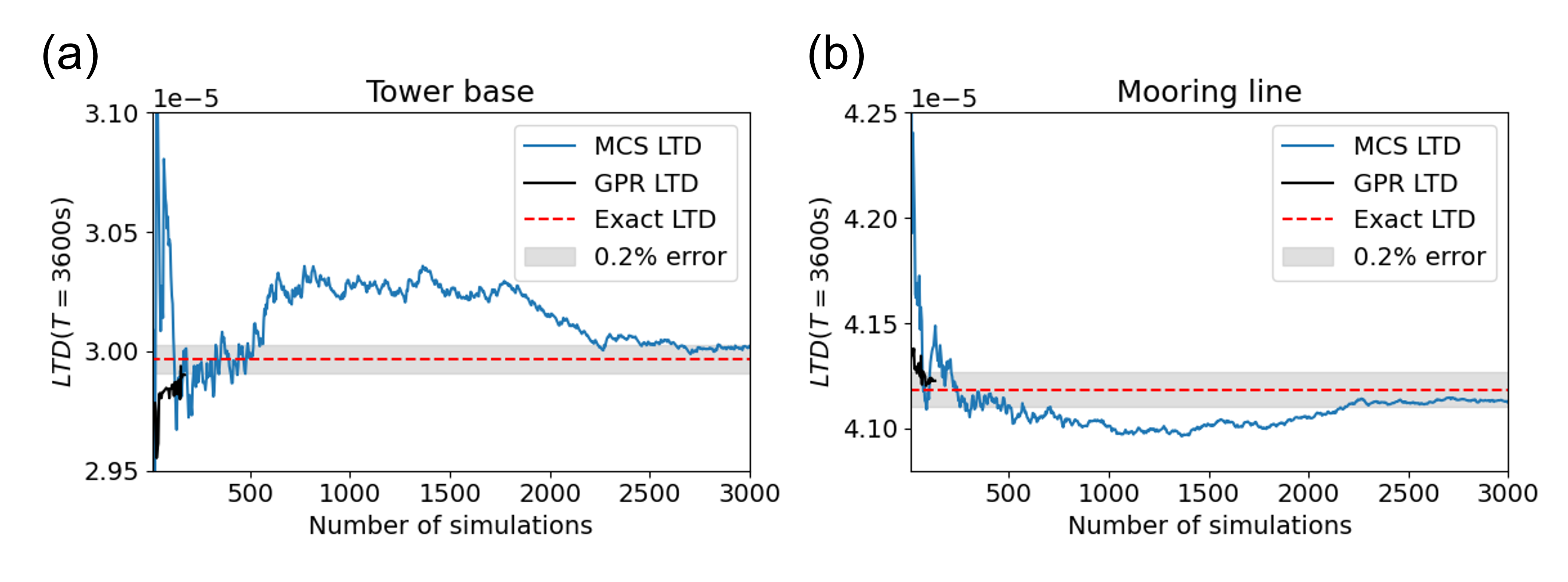}
\caption{Comparison in convergence rates for $LTD(T=\SI{3600}{\second})$ estimation between MCS and GPR approaches for (a) the tower base bending moment; and (b) the mooring line fairlead tension.\label{fig:converge}}
\end{figure}

Lastly, improvement towards final $LTD$ $(T=\SI{3600}{\second})$ estimates is presented in Fig.~\ref{fig:converge}. We compare the convergence rate of our proposed approach with Monte Carlo simulation in terms of the number of FD simulations. Note that MCS used in Fig.~\ref{long-term} and the one used in this comparison are different. MCS in Fig.~\ref{long-term} was applied with both exact and GPR surfaces to show consistency in the response surface learned by our proposed approach. Here, MCS serves as a naive sea state selecting method that simply draws sea states randomly from the joint distribution to add more FD simulations until $LTD$ convergence. Figure~\ref{fig:converge} shows that our approach that selects the next sea states, which accounts for both joint distribution and GPR model uncertainty, converges more than 10 times faster.
For the tower base, MCS require more than 2,500 FD simulations to converge to within a 0.2\% error; in contrast, the final solution with the proposed GPR-based approach achieves the same accuracy with only 202 simulations. For the mooring line, Monte Carlo simulations needed 2,000 FD simulations to converge to within a 0.2\% error while our GPR approach only needs 162 simulations to converge to within a 0.1\% error. The proposed approach ensures an accurate solution with a much smaller number of simulations. 

In this study, we restricted our attention to fatigue damage computations associated with only two response variables. A computationally tractable FD model provided needed evaluations of motions and stresses.  An efficient GPR surrogate model made it possible to compute long-term fatigue damage for the selected response variables---the tower base bending moment and the tension at the fairlead of a mooring line.  In practical cases involving fatigue limit state evaluation, a system reliability formulation might be needed where multiple failure modes and/or hot-spot stress locations must be evaluated.  This was not done here but the model proposed here can be easily extended to consider it.


\section{Conclusions} 
An efficient approach for fatigue damage assessment of a floating offshore wind turbine was proposed in this study. A Gaussian process regression-based surrogate model for seastate-dependent short-term $DEL_\text{1-Hz}$ was developed by using a coupled FD simulation model of a semi-submersible floating offshore wind turbine. First, we selected representative sea states from the analysis of site-specific metocean data (by separate consideration of different wind speed ranges that span the turbine's operating range) as input conditions for the FD FOWT model. The required output PSDs are used in spectral approaches that allowed estimation of short-term fatigue damage. Then, Gaussian process regression was applied to generate two-dimensional response surfaces of $DEL_\text{1-Hz}$ for wave conditions in each wind speed bin. By considering the predictive variance as well as the probability of occurrence of the different sea states, the GPR surrogate model was iteratively improved by performing additional simulations at sea states with higher uncertainty. 

Through this active learning process, the final converged GPR response surfaces were proven to be accurate compared to the exact solution using full-grid calculation. Long-term fatigue damage was estimated using the response surfaces and the joint probability of the sea states. If we consider the number of simulations to be run as an indicator of computational efficiency, the proposed approach appears to be significantly more efficient compared to direct Monte Carlo simulations without a surrogate. The convergence rate of the proposed approach (with only 0.2\% prediction error) was around 10 times faster than with direct Monte Carlo simulations. Finally, we close by noting that the use of a GPR-based surrogate model for improved efficiency in fatigue studies was demonstrated for a relatively smaller wind turbine of 5-MW rating than might be considered more common today. The approach outlined here is not, however, limited to any specific turbine or size; it can be readily adopted for larger wind turbines (of 10-15MW rating) that are being increasingly used offshore. \textcolor{black}{In this study, the efficiency of a GPR-based surrogate model and an active learning process has been validated by using a coupled FD model. This approach can meet the rising need for reliability-based FOWT design optimization, which could lower the levelized cost of clean energy through the development of safer and more reliable floating platforms. Additionally, future studies could extend the application of the proposed method for use with time-domain simulations and high-fidelity computational fluid dynamics simulations that include complex aerodynamic phenomena in more detailed FOWT design studies.}


\section*{Acknowledgments}
EM was partially supported by the PRIN 2022 project “NonlinEar Phenomena in floaTing offshore wind tUrbiNEs (NEPTUNE)”, prot. 2022W7SKTL, funded by the Italian MUR. This support is gratefully acknowledged.

 \bibliographystyle{elsarticle-num} 
 \bibliography{cas-refs}

\begin{thebibliography}{10}
\expandafter\ifx\csname url\endcsname\relax
  \def\url#1{\texttt{#1}}\fi
\expandafter\ifx\csname urlprefix\endcsname\relax\def\urlprefix{URL }\fi
\expandafter\ifx\csname href\endcsname\relax
  \def\href#1#2{#2} \def\path#1{#1}\fi

\bibitem{Xu_RSER2020}
W.~Xu, Y.~Liu, W.~Wu, Y.~Dong, W.~Lu, Y.~Liu, B.~Zhao, H.~Li, R.~Yang,
  \href{https://www.sciencedirect.com/science/article/pii/S1364032120304585}{Proliferation
  of offshore wind farms in the {N}orth {S}ea and surrounding waters revealed
  by satellite image time series}, Renewable and Sustainable Energy Reviews 133
  (2020) 110167.
\newblock \href {https://doi.org/https://doi.org/10.1016/j.rser.2020.110167}
  {\path{doi:https://doi.org/10.1016/j.rser.2020.110167}}.
\newline\urlprefix\url{https://www.sciencedirect.com/science/article/pii/S1364032120304585}

\bibitem{ThiesRE}
P.~R. Thies, L.~Johanning, V.~Harnois, H.~C. Smith, D.~N. Parish,
  \href{https://www.sciencedirect.com/science/article/pii/S0960148113004667}{Mooring
  line fatigue damage evaluation for floating marine energy converters: Field
  measurements and prediction}, Renewable Energy 63 (2014) 133--144.
\newblock \href {https://doi.org/https://doi.org/10.1016/j.renene.2013.08.050}
  {\path{doi:https://doi.org/10.1016/j.renene.2013.08.050}}.
\newline\urlprefix\url{https://www.sciencedirect.com/science/article/pii/S0960148113004667}

\bibitem{HsuMS}
W.~T. Hsu, K.~P. Thiagarajan, L.~Manuel,
  \href{https://www.sciencedirect.com/science/article/pii/S0951833917300886}{Extreme
  mooring tensions due to snap loads on a floating offshore wind turbine
  system}, Marine Structures 55 (2017) 182--199.
\newblock \href
  {https://doi.org/https://doi.org/10.1016/j.marstruc.2017.05.005}
  {\path{doi:https://doi.org/10.1016/j.marstruc.2017.05.005}}.
\newline\urlprefix\url{https://www.sciencedirect.com/science/article/pii/S0951833917300886}

\bibitem{Liu2108JSEE}
J.~Liu, L.~Manuel, \href{https://doi.org/10.1115/1.4039984}{{Alternative
  Mooring Systems for a Very Large Offshore Wind Turbine Supported by a
  Semisubmersible Floating Platform}}, Journal of Solar Energy Engineering
  140~(5) (2018) 051003.
\newblock \href
  {http://arxiv.org/abs/https://asmedigitalcollection.asme.org/solarenergyengineering/article-pdf/140/5/051003/6378951/sol\_140\_05\_051003.pdf}
  {\path{arXiv:https://asmedigitalcollection.asme.org/solarenergyengineering/article-pdf/140/5/051003/6378951/sol\_140\_05\_051003.pdf}},
  \href {https://doi.org/10.1115/1.4039984} {\path{doi:10.1115/1.4039984}}.
\newline\urlprefix\url{https://doi.org/10.1115/1.4039984}

\bibitem{manuel2018alternative}
L.~Manuel, P.~T. Nguyen, J.~Canning, R.~G. Coe, A.~C. Eckert-Gallup, N.~Martin,
  Alternative approaches to develop environmental contours from metocean data,
  Jnl.\ of Ocean Engineering \& Marine Energy 4~(4) (2018) 293--310.

\bibitem{MarinoRE}
E.~Marino, A.~Giusti, L.~Manuel,
  \href{https://www.sciencedirect.com/science/article/pii/S0960148116308886}{Offshore
  wind turbine fatigue loads: The influence of alternative wave modeling for
  different turbulent and mean winds}, Renewable Energy 102 (2017) 157--169.
\newblock \href {https://doi.org/https://doi.org/10.1016/j.renene.2016.10.023}
  {\path{doi:https://doi.org/10.1016/j.renene.2016.10.023}}.
\newline\urlprefix\url{https://www.sciencedirect.com/science/article/pii/S0960148116308886}

\bibitem{ZieglerEP}
L.~Ziegler, S.~Voormeeren, S.~Schafhirt, M.~Muskulus,
  \href{https://www.sciencedirect.com/science/article/pii/S1876610215021542}{Sensitivity
  of wave fatigue loads on offshore wind turbines under varying site
  conditions}, Energy Procedia 80 (2015) 193--200.
\newblock \href {https://doi.org/https://doi.org/10.1016/j.egypro.2015.11.422}
  {\path{doi:https://doi.org/10.1016/j.egypro.2015.11.422}}.
\newline\urlprefix\url{https://www.sciencedirect.com/science/article/pii/S1876610215021542}

\bibitem{JWEIA2008Korn}
K.~Saranyasoontorn, L.~Manuel,
  \href{https://www.sciencedirect.com/science/article/pii/S0167610508000068}{On
  the propagation of uncertainty in inflow turbulence to wind turbine loads},
  Journal of Wind Engineering and Industrial Aerodynamics 96~(5) (2008)
  503--523.
\newblock \href {https://doi.org/https://doi.org/10.1016/j.jweia.2008.01.005}
  {\path{doi:https://doi.org/10.1016/j.jweia.2008.01.005}}.
\newline\urlprefix\url{https://www.sciencedirect.com/science/article/pii/S0167610508000068}

\bibitem{jonkman2011dynamics}
J.~M. Jonkman, D.~Matha, Dynamics of offshore floating wind turbines—analysis
  of three concepts, Wind Energy 14~(4) (2011) 557--569.

\bibitem{liu2017establishing}
Y.~Liu, Q.~Xiao, A.~Incecik, C.~Peyrard, D.~Wan, Establishing a fully coupled
  {CFD} analysis tool for floating offshore wind turbines, Renewable Energy 112
  (2017) 280--301.

\bibitem{kvittem2015frequency}
M.~I. Kvittem, T.~Moan, Frequency versus time domain fatigue analysis of a
  semisubmersible wind turbine tower, Journal of Offshore Mechanics and Arctic
  Engineering 137~(1) (2015).

\bibitem{karimi2017multi}
M.~Karimi, M.~Hall, B.~Buckham, C.~Crawford, A multi-objective design
  optimization approach for floating offshore wind turbine support structures,
  Journal of Ocean Engineering and Marine Energy 3 (2017) 69--87.

\bibitem{karimi2019fully}
M.~Karimi, B.~Buckham, C.~Crawford, A fully coupled frequency domain model for
  floating offshore wind turbines, Journal of Ocean Engineering and Marine
  Energy 5 (2019) 135--158.

\bibitem{Ferri2022}
G.~Ferri, E.~Marino, N.~Bruschi, C.~Borri, {Platform and mooring system
  optimization of a 10 MW semisubmersible offshore wind turbine}, Renewable
  Energy 182 (2022) 1152--1170.
\newblock \href {https://doi.org/10.1016/j.renene.2021.10.060}
  {\path{doi:10.1016/j.renene.2021.10.060}}.

\bibitem{Ferri2023}
G.~Ferri, E.~Marino,
  \href{https://doi.org/10.1016/j.renene.2022.11.116}{{Site-specific
  optimizations of a 10 MW floating offshore wind turbine for the Mediterranean
  Sea}}, Renewable Energy 202~(November 2022) (2023) 921--941.
\newblock \href {https://doi.org/10.1016/j.renene.2022.11.116}
  {\path{doi:10.1016/j.renene.2022.11.116}}.
\newline\urlprefix\url{https://doi.org/10.1016/j.renene.2022.11.116}

\bibitem{tran2015aerodynamic}
T.~T. Tran, D.~H. Kim, The aerodynamic interference effects of a floating
  offshore wind turbine experiencing platform pitching and yawing motions,
  Journal of Mechanical Science and Technology 29 (2015) 549--561.

\bibitem{kyle2020propeller}
R.~Kyle, Y.~C. Lee, W.-G. Fr{\"u}h, Propeller and vortex ring state for
  floating offshore wind turbines during surge, Renewable Energy 155 (2020)
  645--657.

\bibitem{cai2023effects}
Y.~Cai, H.~Zhao, X.~Li, Y.~Liu, Effects of yawed inflow and blade-tower
  interaction on the aerodynamic and wake characteristics of a horizontal-axis
  wind turbine, Energy 264 (2023) 126246.

\bibitem{dose2020fluid}
B.~Dose, H.~Rahimi, B.~Stoevesandt, J.~Peinke, Fluid-structure coupled
  investigations of the nrel 5 mw wind turbine for two downwind configurations,
  Renewable energy 146 (2020) 1113--1123.

\bibitem{santo2020effect}
G.~Santo, M.~Peeters, W.~Van~Paepegem, J.~Degroote, Effect of rotor--tower
  interaction, tilt angle, and yaw misalignment on the aeroelasticity of a
  large horizontal axis wind turbine with composite blades, Wind Energy 23~(7)
  (2020) 1578--1595.

\bibitem{zheng2023efficient}
S.~Zheng, C.~Li, Y.~Xiao, Efficient optimization design method of jacket
  structures for offshore wind turbines, Marine Structures 89 (2023) 103372.

\bibitem{wilson2017surrogate}
B.~Wilson, S.~Wakes, M.~Mayo, Surrogate modeling a computational fluid
  dynamics-based wind turbine wake simulation using machine learning, in: 2017
  IEEE Symposium Series on Computational Intelligence (SSCI), IEEE, 2017, pp.
  1--8.

\bibitem{zahle2018computational}
F.~Zahle, N.~N. S{\o}rensen, M.~K. McWilliam, A.~Barlas, Computational fluid
  dynamics-based surrogate optimization of a wind turbine blade tip extension
  for maximising energy production, in: Journal of Physics: Conference Series,
  Vol. 1037, IOP Publishing, 2018, p. 042013.

\bibitem{lim2021RESS}
H.~Lim, L.~Manuel, Distribution-free polynomial chaos expansion surrogate
  models for efficient structural reliability analysis, Reliability Engineering
  \& System Safety 205 (2021) 107256.
\newblock \href {https://doi.org/https://doi.org/10.1016/j.ress.2020.107256}
  {\path{doi:https://doi.org/10.1016/j.ress.2020.107256}}.

\bibitem{li2019wind}
Y.~Li, S.~Liu, L.~Shu, Wind turbine fault diagnosis based on {G}aussian process
  classifiers applied to operational data, Renewable Energy 134 (2019)
  357--366.

\bibitem{avendano2021virtual}
L.~D. Avenda{\~n}o-Valencia, I.~Abdallah, E.~Chatzi, Virtual fatigue
  diagnostics of wake-affected wind turbine via {G}aussian process regression,
  Renewable Energy 170 (2021) 539--561.

\bibitem{zhang2022hybrid}
S.~Zhang, E.~Robinson, M.~Basu, Hybrid {G}aussian process regression and fuzzy
  inference system based approach for condition monitoring at the rotor side of
  a doubly fed induction generator, Renewable Energy 198 (2022) 936--946.

\bibitem{abdallah2019parametric}
I.~Abdallah, C.~Lataniotis, B.~Sudret, Parametric hierarchical kriging for
  multi-fidelity aero-servo-elastic simulators—application to extreme loads
  on wind turbines, Probabilistic Engineering Mechanics 55 (2019) 67--77.

\bibitem{taflanidis2013offshore}
A.~A. Taflanidis, E.~Loukogeorgaki, D.~C. Angelides, Offshore wind turbine risk
  quantification/evaluation under extreme environmental conditions, Reliability
  Engineering \& System Safety 115 (2013) 19--32.

\bibitem{lim2021JOMAE}
H.~Lim, L.~Manuel, Y.~Min~Low, On efficient surrogate model development for the
  prediction of the long-term extreme response of a moored floating structure,
  Journal of Offshore Mechanics and Arctic Engineering 143~(1) (2021) 011703.
\newblock \href {https://doi.org/https://doi.org/10.1115/1.4047545}
  {\path{doi:https://doi.org/10.1115/1.4047545}}.

\bibitem{leimeister2018review}
M.~Leimeister, A.~Kolios, A review of reliability-based methods for risk
  analysis and their application in the offshore wind industry, Renewable and
  Sustainable Energy Reviews 91 (2018) 1065--1076.

\bibitem{wilkie2021gaussian}
D.~Wilkie, C.~Galasso, {G}aussian process regression for fatigue reliability
  analysis of offshore wind turbines, Structural Safety 88 (2021) 102020.

\bibitem{pillai2018mooring}
A.~C. Pillai, P.~R. Thies, L.~Johanning, Mooring system design optimization
  using a surrogate assisted multi-objective genetic algorithm, Engineering
  Optimization (2018).

\bibitem{Lim2022ES}
H.~Lim, L.~Manuel, Y.~M. Low, N.~Srinil,
  \href{https://www.sciencedirect.com/science/article/pii/S0141029621018678}{A
  surrogate model for estimating uncertainty in marine riser fatigue damage
  resulting from vortex-induced vibration}, Engineering Structures 254 (2022)
  113796.
\newblock \href
  {https://doi.org/https://doi.org/10.1016/j.engstruct.2021.113796}
  {\path{doi:https://doi.org/10.1016/j.engstruct.2021.113796}}.
\newline\urlprefix\url{https://www.sciencedirect.com/science/article/pii/S0141029621018678}

\bibitem{singh2022probabilistic}
D.~Singh, R.~P. Dwight, K.~Laugesen, L.~Beaudet, A.~Vir{\'e}, Probabilistic
  surrogate modeling of offshore wind-turbine loads with chained {G}aussian
  processes, in: Journal of Physics: Conference Series, Vol. 2265, IOP
  Publishing, 2022, p. 032070.

\bibitem{teixeira2017analysis}
R.~Teixeira, A.~O’Connor, M.~Nogal, N.~Krishnan, J.~Nichols, Analysis of the
  design of experiments of offshore wind turbine fatigue reliability design
  with kriging surfaces, Procedia Structural Integrity 5 (2017) 951--958.

\bibitem{murcia2018uncertainty}
J.~P. Murcia, P.~E. R{\'e}thor{\'e}, N.~Dimitrov, A.~Natarajan, J.~D.
  S{\o}rensen, P.~Graf, T.~Kim, Uncertainty propagation through an aeroelastic
  wind turbine model using polynomial surrogates, Renewable Energy 119 (2018)
  910--922.

\bibitem{shi2019AOR}
C.~Shi, L.~Manuel, Non-parametric prediction of the long-term fatigue damage
  for an instrumented top-tensioned riser, Applied Ocean Research 82 (2019)
  245--258.
\newblock \href {https://doi.org/https://doi.org/10.1016/j.apor.2018.11.001}
  {\path{doi:https://doi.org/10.1016/j.apor.2018.11.001}}.

\bibitem{muller2018application}
K.~M{\"u}ller, P.~W. Cheng, Application of a {M}onte {C}arlo procedure for
  probabilistic fatigue design of floating offshore wind turbines, Wind Energy
  Science 3~(1) (2018) 149--162.

\bibitem{richmond2020stochastic}
M.~Richmond, A.~Sobey, R.~Pandit, A.~Kolios, Stochastic assessment of
  aerodynamics within offshore wind farms based on machine-learning, Renewable
  Energy 161 (2020) 650--661.

\bibitem{gasparis2020surrogate}
G.~Gasparis, W.~H. Lio, F.~Meng, Surrogate models for wind turbine electrical
  power and fatigue loads in wind farm, Energies 13~(23) (2020) 6360.

\bibitem{li2020long}
X.~Li, W.~Zhang, Long-term fatigue damage assessment for a floating offshore
  wind turbine under realistic environmental conditions, Renewable Energy 159
  (2020) 570--584.

\bibitem{dimitrov2018wind}
N.~Dimitrov, M.~C. Kelly, A.~Vignaroli, J.~Berg, From wind to loads: wind
  turbine site-specific load estimation with surrogate models trained on
  high-fidelity load databases, Wind Energy Science 3~(2) (2018) 767--790.

\bibitem{ankenman2008stochastic}
B.~Ankenman, B.~L. Nelson, J.~Staum, Stochastic kriging for simulation
  metamodeling, in: 2008 Winter simulation conference, IEEE, 2008, pp.
  362--370.

\bibitem{yue2018surrogate}
X.~Yue, Y.~Wen, J.~H. Hunt, J.~Shi, Surrogate model-based control considering
  uncertainties for composite fuselage assembly, Journal of Manufacturing
  Science and Engineering 140~(4) (2018).

\bibitem{wang2017robust}
D.~Wang, J.~Xue, D.~Cui, Y.~Zhong, A robust submap-based road shape estimation
  via iterative {G}aussian process regression, in: 2017 IEEE Intelligent
  Vehicles Symposium (IV), IEEE, 2017, pp. 1776--1781.

\bibitem{li2021robust}
Z.-Z. Li, L.~Li, Z.~Shao, Robust {G}aussian process regression based on
  iterative trimming, Astronomy and Computing 36 (2021) 100483.

\bibitem{pasolli2011gaussian}
E.~Pasolli, F.~Melgani, {G}aussian process regression within an active learning
  scheme, in: 2011 IEEE International Geoscience and Remote Sensing Symposium,
  IEEE, 2011, pp. 3574--3577.

\bibitem{yue2020active}
X.~Yue, Y.~Wen, J.~H. Hunt, J.~Shi, Active learning for {G}aussian process
  considering uncertainties with application to shape control of composite
  fuselage, IEEE Transactions on Automation Science and Engineering 18~(1)
  (2020) 36--46.

\bibitem{lam2008sequential}
C.~Q. Lam, Sequential adaptive designs in computer experiments for response
  surface model fit, Ph.D. thesis, The Ohio State University (2008).

\bibitem{yan2017high}
H.~Yan, High dimensional data analysis for anomaly detection and quality
  improvement, Ph.D. thesis, Georgia Institute of Technology (2017).

\bibitem{kowalska2012maritime}
K.~Kowalska, L.~Peel, Maritime anomaly detection using {G}aussian process
  active learning, in: 2012 15th International Conference on Information
  Fusion, IEEE, 2012, pp. 1164--1171.

\bibitem{stieng2020reliability}
L.~E.~S. Stieng, M.~Muskulus, Reliability-based design optimization of offshore
  wind turbine support structures using analytical sensitivities and factorized
  uncertainty modeling, Wind Energy Science 5~(1) (2020) 171--198.

\bibitem{oc42014}
A.~Robertson, J.~Jonkman, {Definition of the Semisubmersible Floating System
  for Phase II of OC4}, Tech. Rep. September, National Renewable Energy
  Laboratory Golden, CO, USA (2014).

\bibitem{Borgman1969}
L.~E. Borgman, {Ocean Wave Simulation for Engineering Design}, Journal of the
  Waterways and Harbors Division 95~(4) (1969) 557--583.
\newblock \href {https://doi.org/10.1061/jwheau.0000665}
  {\path{doi:10.1061/jwheau.0000665}}.

\bibitem{Jonkman2007}
J.~Jonkman, M.~Buhl, \href{http://www.ncbi.nlm.nih.gov/pubmed/21564034}{{FAST
  User's Guide}}, Tech. Rep.~6, National Renewable Energy Laboratory Golden,
  CO, USA (2007).
\newblock \href {http://arxiv.org/abs/ArXiv ID} {\path{arXiv:ArXiv ID}}, \href
  {https://doi.org/10.2172/15020796} {\path{doi:10.2172/15020796}}.
\newline\urlprefix\url{http://www.ncbi.nlm.nih.gov/pubmed/21564034}

\bibitem{Bir2008}
G.~S. Bir, \href{http://wind.nrel.gov/designcodes/postprocessors/mbc}{User’s
  guide to {MBC3} (multi-blade coordinate transformation utility for 3-bladed
  wind turbines)}, Tech. Rep. October, National Renewable Energy Laboratory
  Golden, CO, USA (2008).
\newline\urlprefix\url{http://wind.nrel.gov/designcodes/postprocessors/mbc}

\bibitem{scott2015multivariate}
D.~W. Scott, Multivariate density estimation: theory, practice, and
  visualization, John Wiley \& Sons, 2015.

\bibitem{dirlik1985application}
T.~Dirlik, Application of computers in fatigue analysis, Ph.D. thesis, PhD
  Thesis, University of Warwick (1985).

\bibitem{Dirlik2021Metals}
T.~Dirlik, D.~Benasciutti, Dirlik and {T}ovo-{B}enasciutti spectral methods in
  vibration fatigue: a review with a historical perspective, Metals 11~(9)
  (2021) 1333.

\bibitem{Ragan2007WindEngineering}
P.~Ragan, L.~Manuel,
  \href{https://doi.org/10.1260/030952407781494494}{Comparing estimates of wind
  turbine fatigue loads using time-domain and spectral methods}, Wind
  Engineering 31~(2) (2007) 83--99.
\newblock \href
  {http://arxiv.org/abs/https://doi.org/10.1260/030952407781494494}
  {\path{arXiv:https://doi.org/10.1260/030952407781494494}}, \href
  {https://doi.org/10.1260/030952407781494494}
  {\path{doi:10.1260/030952407781494494}}.
\newline\urlprefix\url{https://doi.org/10.1260/030952407781494494}

\bibitem{williams1995gaussian}
C.~Williams, C.~Rasmussen, {G}aussian processes for regression, Advances in
  neural information processing systems 8 (1995).

\bibitem{RobertsPTSA}
S.~Roberts, M.~Osborne, M.~Ebden, S.~Reece, N.~Gibson, S.~Aigrain,
  \href{https://royalsocietypublishing.org/doi/abs/10.1098/rsta.2011.0550}{{G}aussian
  processes for time-series modelling}, Phil.\ Trans.\ of the Royal Society A:
  Math., Phy.\ and Eng.\ Sciences 371~(1984) (2013) 20110550.
\newblock \href
  {http://arxiv.org/abs/https://royalsocietypublishing.org/doi/pdf/10.1098/rsta.2011.0550}
  {\path{arXiv:https://royalsocietypublishing.org/doi/pdf/10.1098/rsta.2011.0550}},
  \href {https://doi.org/10.1098/rsta.2011.0550}
  {\path{doi:10.1098/rsta.2011.0550}}.
\newline\urlprefix\url{https://royalsocietypublishing.org/doi/abs/10.1098/rsta.2011.0550}

\bibitem{gardner2018gpytorch}
J.~Gardner, G.~Pleiss, K.~Q. Weinberger, D.~Bindel, A.~G. Wilson, G{P}y{T}orch:
  Blackbox matrix-matrix {G}aussian process inference with gpu acceleration,
  Advances in neural information processing systems 31 (2018).

\bibitem{dnv2016dnvgl}
{Det Norske Veritas}, {DNVGL-RP-C203}: Fatigue design of offshore steel
  structures, DNV, Oslo, Norway (2016).

\end{thebibliography}





\end{document}